\documentclass[preprint,preprintnumbers,aps,superscriptaddress,nofootinbib,tightenlines,floatfix]{revtex4}

\usepackage{amsmath,amssymb,bm}
\usepackage{subfigure}
\usepackage{verbatim}

\newcommand{\beq}{\begin{equation}}
\newcommand{\eeq}{\end{equation}}
\newcommand{\bea}{\begin{eqnarray}}
\newcommand{\eea}{\end{eqnarray}}

\newcommand{\lsim}{\raisebox{-0.7ex}{$\stackrel{\textstyle <}{\sim}$ }}

\ifx\pdfoutput\undefined
\usepackage[dvips]{graphicx}
\else
\usepackage[pdftex]{graphicx}
\fi
\usepackage{epstopdf}

\begin{document}
\DeclareGraphicsExtensions{.pdf,.gif,.jpg}

\preprint{\vbox{ 
\hbox{NT@UW-10-14} 
}}

\title{Nucleon-Nucleon Scattering in a Harmonic Potential}

\author{Thomas Luu}
\email[]{tluu@llnl.gov}
\affiliation{N
  Section, Lawrence Livermore National Laboratory, Livermore, CA
  94551, USA}
\author{Martin Savage}
\email[]{mjs5@u.washington.edu}
\affiliation{Department of Physics, University of Washington, Seattle,
  WA 98195-1560, USA}  
\author{Achim Schwenk}
\email[]{schwenk@physik.tu-darmstadt.de}
\affiliation{ExtreMe Matter Institute EMMI, GSI Helmholtzzentrum f\"ur
  Schwerionenforschung GmbH, 64291 Darmstadt, Germany}
\affiliation{Institut f\"ur Kernphysik, Technische Universit\"at Darmstadt, 64289 Darmstadt, Germany}
\affiliation{TRIUMF, 4004 Wesbrook Mall, Vancouver BC, V6T 2A3, Canada}
\author{James P. Vary}
\email[]{jvary@iastate.edu}
\affiliation{Department of Physics and Astronomy, Iowa State University, 
Ames, IA 50011, USA}

\date{\today}

\begin{abstract}
The discrete energy-eigenvalues of two nucleons interacting with a finite-range nuclear force and confined
to a harmonic potential are used to numerically reconstruct the
free-space scattering phase shifts. The extracted phase shifts are
compared to those obtained from the exact continuum scattering
solution and agree within the uncertainties of the calculations. Our
results suggest that it might be possible to determine the
amplitudes for the scattering of complex systems, such as $nd$, $nt$
or $n\alpha$, from the energy-eigenvalues confined to finite volumes
using \emph{ab-initio} bound-state techniques.
\end{abstract}


\maketitle

\section{Introduction}
\label{sect:intro}

\noindent
Quantum scattering of strongly interacting few-nucleon systems is
complicated and requires careful treatment of asymptotics,
antisymmetrization effects as well as the dynamics generated by
nuclear forces. Full treatments of antisymmetrization with
correlations have become routine in bound-state and quasi-bound-state
solutions of light nuclei using \emph{ab-initio} techniques based on
nucleon-nucleon ($NN$) and three-nucleon interactions ($3N$).
Techniques, such as the No Core Shell Model (NCSM) (see, e.g.,
Ref.~\cite{Navratil:2000gs,Navratil:2007we}), Green's Function Monte
Carlo (GFMC) (see, e.g., Ref.~\cite{Pieper:2007ax,Pieper:2004qw}), and
the Coupled Cluster (CC) approach (see, e.g.,
Ref.~\cite{Hagen:2008iw,Hagen:2007ew}), are used to calculate
ground and excited states of light nuclei. The precision of these
calculations has reached a point where further progress is now limited
by the fidelity of the input interactions.

In reaction calculations of scattering properties of light nuclei,
progress has been less pronounced, though nonetheless significant.
R-matrix analysis (see, e.g., Ref.~\cite{Lane:1958,Descouvemont:2010cx}) has been
historically the empirical workhorse, providing impressive fits to a range of experimental data.    More microscopic approaches to the scattering of light nuclei are based on the Resonating Group Method (RGM)~\cite{Langanke:1986,Fujimura:1999zz,Pfitzinger:2000nw}. A promising avenue for \emph{ab-initio} calculations of scattering of light ions
comes from the coupling of the RGM reaction method with the NCSM bound-state
technique~\cite{Quaglioni:2008sm}.  For this method, the large computational resources required to achieve convergence
provide the limiting constraint on reliably calculating scattering
parameters for processes with $A > 5$.  It would be significant
if existing bound-state techniques, and their accompanying precision,
could be further exploited to reliably determine scattering amplitudes
for multi-nucleon systems.

During the last twenty years the general technique of 
effective field theory (EFT) has been developed and applied to multi-nucleon
systems.
Effective Field Theory provides a description of observables,
consistent with the approximate chiral symmetries of quantum chromodynamics (QCD),
in terms of a small number of expansion parameters within a plane-wave basis.  These expansion parameters are used to relate the experimentally determined scattering
parameters and bound-state properties of few-nucleon systems
to the coefficients of operators at a given order in the EFT expansion.  In principle, this 
allows for systematically improvable 
calculations of multi-nucleon observables.
Applying the EFT framework within an oscillator basis has also been investigated~\cite{Haxton:2002kb,Stetcu:2006ey,Haxton:2007hx}.
More recently, it has been suggested that the EFT framework might be fruitfully
applied to multi-fermion systems confined in a harmonic 
potential~\cite{Luu:2006xv,Stetcu:2007ms,Stetcu:2009ic}, and might be
usefully married with the NCSM calculational scheme.  In Ref.~\cite{bira:exascale} it was suggested that the scattering properties of certain complex nuclear systems could be calculated from the spectrum of the same systems confined to a harmonic potential.  This was demonstrated for two confined particles at the unitary limit in Ref.~\cite{Stetcu:2010xq}.  A lattice formulation of EFT coupled with an external harmonic potential is currently being developed~\cite{amy}.

In this work we investigate the simple two-nucleon system
confined in a harmonic potential of the form $V_{HO}={1\over 2}M_N \omega^2 r^2$,
and interacting via nuclear forces in uncoupled partial waves.  
Since two-body techniques are well established for both scattering and bound 
states, this system is ideal for determining the extent to which continuum scattering
amplitudes can be recovered from bound-state information.
An analytic expression that relates the eigenvalues of two interacting
particles moving in a harmonic potential to the scattering phase shift at those energies,
analogous to 
``L\"uschers method''~\cite{Hamber:1983vu,Luscher:1986pf,Luscher:1990ux}
that is used in Lattice QCD, 
allows for the scattering phase shift to be determined in the limit
that the oscillator length is large compared to the range of nuclear forces.
As the latter is characterized by the Compton wavelength of the pion,
with increasing confinement this leads to modifications to nuclear
forces due to the harmonic potential that
must be systematically removed in order to accurately predict
the scattering amplitude.
This is achieved
by calculating the energy-eigenvalues over a range of harmonic
frequencies, $\omega$, 
determining the scattering amplitude
over this range of $\omega$,
and then extrapolating to the $\omega=0$ limit.
We verify the extracted phase shifts by comparing them to the results
of an independent scattering calculation.
The two methods are found to yield the same phase shifts  within the uncertainties of
the calculations.

\section{Phase Shifts from the  Zero-Range Relation}

\noindent
It is well-established that the energy-eigenvalues of an interacting system 
of particles 
confined to a finite volume can be used to determine the scattering
phase shift
(at the energy-eigenvalues) when the size of the confining region is much larger 
than the range of the interactions between the particles.  
For instance, relating the scattering phase shift to the energy-eigenvalues of 
two-nucleons confined to a spherical region by solving the Schr$\ddot{\text{o}}$dinger equation
with a Dirichlet
boundary condition is a problem that appears in standard texts on nuclear
physics, see, e.g., Ref.~\cite{PrestonBhadhuri}.  
This method has been successfully employed in the latticization of
low-energy EFT's to predict (to a given level of precision)
the ground state energies of light nuclei and their volume dependence~\cite{Epelbaum:2010xt}.

Volume dependence is also used to determine meson-meson,
meson-baryon and baryon-baryon elastic scattering phase shifts from the
energy-eigenvalues of these systems calculated with 
Lattice QCD~\footnote{The Maiani-Testa theorem~\cite{Maiani:1990ca}
precludes the extraction of
  scattering matrix elements from Euclidean-space Green functions in the
  infinite-volume limit except at kinematic thresholds.}
(for a recent review, see Ref.~\cite{Beane:2010em}).  
Lattice QCD calculations are generally performed in spatial volumes with cubic
symmetry and with periodic boundary-conditions (BC's) imposed upon the fields at the
edges.
This reduces the number of momentum modes, and hence reduces
the kinetic-energy contributions to the calculated processes.
The
non-relativistic relation
between the energy-eigenvalues and the scattering amplitude 
(the ``L\"uscher relation'')
has been shown to be valid
even in quantum field theory~\cite{Luscher:1986pf,Luscher:1990ux}.
Given the energy-splitting, 
$\Delta E_n$, between the two-hadron state and the hadron masses,
$m$,  the real part of the inverse scattering amplitude below inelastic
thresholds is 
\begin{eqnarray}
p_n \cot \delta(p_n) \ =\ {1\over \pi L}\ {\bf
  S}\left(\,\left(\frac{p_n L}{2\pi}\right)^2\,\right)
\ \ ,\ \ 
{\bf S}\left(\, x \, \right)\ \equiv \ \sum_{\bf j}^{ |{\bf j}|<\Lambda}
{1\over |{\bf j}|^2-x}\ -\  {4 \pi \Lambda}
\ \ ,
\label{eq:LuscherRelation}
\end{eqnarray}
%
where $\Delta E_n \ =\ \ 2\sqrt{\ p_n^2\ +\ m^2\ } \ -\ 2m$, $\delta(p_n)$ is the energy-dependent 
phase-shift, and the
limit $\Lambda\rightarrow\infty$ is implicit.  The ${\bf S}$-function is the
Green function, $G_{HH}({\bf 0},{\bf 0})$, 
for two-hadron   plane-wave eigenstates (and straightforwardly generalizes to
hadrons with different masses).
The L\"uscher relation 
between the scattering amplitude and the 
energy-eigenvalues in the
finite lattice volume,
given by Eq.~(\ref{eq:LuscherRelation}),
is valid when the spatial extent of the lattice is
large compared to the range of the interaction, $R$.  
Corrections to the relation
are found to behave as $\sim e^{-L/R}$, see, e.g., Ref.~\cite{Bedaque:2006yi}.
If, in the continuum (infinite-volume) limit, a system contains a shallow
bound state, as is the case in the NN $^3S_1$-$^3D_1$ coupled
channel, the periodic BC in a finite volume increases the binding energy of the
state. The finite-volume corrections scale as $e^{-\gamma_0 L}$~\cite{Beane:2003da}, where
$\gamma_0$ is the binding momentum in the continuum.  In contrast, the
continuum scattering states have power-law dependences upon the lattice extent
for $L\gg R$, with energies that behave as $\sim 1/L^3$ for the ground
state
and $\sim 1/L^2$ for the higher-energy states.

The systems that we considered in this work are comprised of two
nucleons in a harmonic potential (with oscillator frequency $\omega$)
interacting through NN
forces. 
We will consider the JISP16 potential~\cite{Shirokov:2005bk,JISP16_web,Maris:2008ax}, which reproduces the low-energy NN scattering data with a $\chi^2/{\rm dof}\sim1.0$, but our results are general and the technique can be applied to other NN interactions.
The EFT-method that was used to (re-)derive the L\"uscher relation 
in Eq.~(\ref{eq:LuscherRelation}) in  non-relativistic quantum 
mechanics~\cite{Beane:2003da}, can
be used to (re-)derive the relation between $p^{2l+1} \cot\delta_l(p)$
in an uncoupled partial wave with angular momentum $l$ and the
energy-eigenvalues of two nucleons in a harmonic 
potential~\cite{Busch:1997,Suzuki:2009,Yip:2008},
\begin{equation}
p^{2l+1}\text{cot}\delta_l(p)\ =\ 
(-1)^{l+1}\left(2 m \omega \right)^{l+1/2}
\frac{\Gamma\left(\frac{2l+3}{4}-\frac{\epsilon}{2}\right)}{\Gamma\left(
\frac{1-2l}{4}-\frac{\epsilon}{2}\right)}
\ \ \ ,
\label{eq:NNHO}
\end{equation}
where 
$\epsilon=E/\omega$ and $E=p^2/m$ is the fully 
interacting energy in the center of mass frame.
While the EFT derivation using the pionless-EFT is valid only up to momenta
associated with the cut in the t-channel from the exchange of one pion, the
relation is valid up to the inelastic threshold.
Equation~(\ref{eq:NNHO}), like the L\"uscher formula,
is valid in the limit of
zero-range interactions.  The harmonic potential,
being non-zero everywhere, except at the origin,  modifies the interaction between
the two nucleons, and the $NN$ phase shift at the outer range of the nuclear potential
differs from that in free space.  This is a finite-range effect, and unlike the
situation encountered in Lattice QCD calculations, it is not expected to be 
exponentially suppressed (in $\omega$).
Equation~(\ref{eq:NNHO}), in conjunction with the leading order (LO)
term in the effective range expansion (ERE) of $p\cot\delta$, the scattering length
$a$, has been used to determine the spectrum of dilute cold atoms in
traps with essentially zero-range interactions, particularly in the
vicinity of Feshbach resonances~\cite{Busch:1997,Mehen:2007dn,Suzuki:2009,Yip:2008}.  
On the other hand, since Eq.~(\ref{eq:NNHO}) relates the
energy-dependent 
phase shift to the energy-eigenvalues $E$ of the confined system, knowledge of the 
spectrum of the two-particle system allows the extraction of the 
continuum scattering amplitude up to finite-range corrections.

\subsection{Loosely Bound States in Weak Harmonic Potentials}
\label{sect:boundstate}

\noindent
For attractive $S$-wave interactions with positive scattering length,
$a_0>0$, a bound state exists and the binding energy $B_0$ can be written in terms of the
binding momentum, $\gamma_0$,
\begin{displaymath}
B_0\equiv\frac{\gamma_0^2}{m}
\ \ \ ,
\end{displaymath}
where $\gamma \sim 1/a$ for scattering lengths large compared to the range of
the interaction.
Refining the estimate of the binding energy gives $\gamma$ 
as the solution to 
\begin{eqnarray}
{1\over a_0} +{1\over 2} r_0 \gamma_0^2 \ -\ \gamma_0\ =\ 0
\ \ \ ,
\end{eqnarray}
where $r_0$ is the effective range of the interaction, and the
ERE of $p\cot\delta = -{1\over a_0}+{1\over 2} r_0 p^2\ +\ ...$
has been truncated at second order, which, in the case of $S$-wave interactions
between nucleons, is sufficient for most purposes.  
The presence of the harmonic potential, and in particular its
non-zero value throughout the volume of the bound-state, gives rise to a
power-law modification to the binding energy~\cite{Busch:1997},
even in the limit of zero-range interactions.
The location of the state corresponding to the free-space bound-state can be
found directly from Eq.~(\ref{eq:NNHO}) in the zero-range limit,
and for small $\omega$ the shift in the
energy of the bound-state is perturbative in $\omega^2$,
\begin{eqnarray}
B_\omega & = & B_0\ -\ \frac{1}{8\left(1-\gamma_0
      r_0\right)}\ \frac{\omega^2}{B_0}\ 
\ +\ O(\omega^4)
\ =\ B_0\ -\ C_{ZR}\ \omega^2+O(\omega^4)
\ \ \ \ , 
\label{eqn:s-waveboundstate}
\end{eqnarray}
where we define
$C_{ZR}=\left[ 8 B_0 \left(1-\gamma_0 r_0\right)\right]^{-1}$ for 
later reference.
Equation~(\ref{eqn:s-waveboundstate}) indicates  that, given the bound-state
energy $B_\omega$
calculated at different values of $\omega$, the continuum binding energy
$B_0$ could be determined by an extrapolation in $\omega^2$ to $\omega=0$.  This same extrapolation can also be done in the presence of finite-range corrections since, as we show later, these corrections occur at order $\omega^2$ for small $\omega$.
In the $^3S_1$-$^3D_1$ coupled-channels that contain the deuteron 
with binding-energy $B_0=2.224575~{\rm MeV}$ ($\gamma_0\sim 45.7~{\rm MeV}$) 
and with an S-wave effective range of $r_0\sim 1.74~{\rm fm}$,  
the coefficient is $C_{ZR}$=0.0944 MeV$^{-1}$.

The LO shift in the bound-state energy 
given in eq.~(\ref{eqn:s-waveboundstate}) 
can be recovered from the bound-state wavefunction based on
ERE,
\begin{eqnarray}
\psi^{(ER)}(r) & = & \psi_{\rm short}(r)\ +\ \sqrt{\gamma_0\over 2\pi}\ 
{1\over\sqrt{1-\gamma_0 r_0}}\
  {e^{-\gamma_0 r}\over r}
\ \ ,
\label{eq:EREpsi}
\end{eqnarray}
where $\psi_{\rm short}(r)$ is the short-distance component of the wavefunction
that has support over a radius $r\ll \gamma_0^{-1}$.
The factor of $1/\sqrt{1-\gamma_0 r_0}$ in eq.~(\ref{eq:EREpsi}) is determined
by the residue of the pole in the scattering amplitude.  
At LO in perturbation theory, the contribution to the energy of this
state from the harmonic potential is
\begin{eqnarray}
\Delta E_0 & =  & 
\langle\psi^{(ER)} | {1\over 2} m \omega^2 r^2 |\psi^{(ER)}\rangle\
\ =\ 
\frac{1}{8\left(1-\gamma_0 r_0\right)}\ \frac{\omega^2}{B_0}\ 
 +\ {\rm short-distance}
\ \ ,
\label{eq:BEshift}
\end{eqnarray}
in agreement with the result in eq.~(\ref{eqn:s-waveboundstate}).

\subsection{Scattering States in Weak Harmonic Potentials}
\label{sect:scatteringstates}
\noindent
It is useful to construct
perturbative expansions for the energy-eigenvalues in the zero-range limit.  
As we show later in sect.~\ref{sect:numericalanalysis}, 
these relations can be used
to readily extract effective range parameters given the low energy 
spectrum of the system.  
Using the zero-range relation given in eq.~(\ref{eq:NNHO}) it is
straightforward to determine the location of the energy-eigenstates in the
limit that $\sqrt{m \omega}/(p\cot\delta)\ll 1$, and also in the unitary-limit where 
$\sqrt{m \omega}/(p\cot\delta)\gg 1$.  In the  $\sqrt{m \omega}/(p\cot\delta)\ll 1$
limit, 
the $q^{th}$ energy-level with orbital angular
momentum $l$ is located at
\begin{eqnarray}
{E_q^{(l)}\over\omega}
& = & 
\left({3\over 2} + l + 2 q\right)
\nonumber\\
&& \ +\ 
2 \left[\ 
\left({\sqrt{2}\over b }\right)^{2l+1}\ 
{(-)^{l+q}\over \Gamma[1+q]\ \Gamma[-{1\over 2}-l-q]\
  p^{2l+1}\cot\delta_l(E_0)}
\right.\nonumber\\
&& \left.
\ \ \ \ \ \ +\ 
\left({\sqrt{2}\over b }\right)^{4l+2}\ 
{ H(-{3\over 2}-l-q) - H(q)
\over
\left[ \ \Gamma[1+q]\ \Gamma[-{1\over 2}-l-q]\
  p^{2l+1}\cot\delta_l(E_0)\ \right]^2}
\ +\ .\ .\ .
\ \right]
\ \ ,
\end{eqnarray}
where
$E_0={1\over m b^2}\left({3\over 2} + l + 2 q\right)$
with $b=1/\sqrt{m\omega}$, and the $H(x)$ are harmonic 
numbers~\footnote{Our definition of the oscillator parameter differs 
from that of Ref.~\cite{Stetcu:2010xq} by a factor of $\sqrt{2}$.}.
When applied to the lowest-lying S-wave state, one finds that, for
small-$\omega$,
\begin{equation}
\label{eqn:s-wave ground state}
\frac{E_0^{(0)}}{\omega}=\frac{3}{2}-\frac{1}{bp\ \mbox{cot}\delta_0}
\left(\sqrt{\frac{2}{\pi}}-\frac{2(1-\log 2)}{\pi\ bp\ \mbox{cot}\delta_0}
-\frac{\pi^2-24+36(2-\log 2)\log 2}{6\sqrt{2}\pi^{3/2}\left(bp\ \mbox{cot}
\delta_0\right)^2}
\ +\ .\ .\ .\right)
\ ,
\end{equation}
where 
$p\ \mbox{cot}\delta_0$ is evaluated at 
$p^2/m=3\omega/2$~\footnote{To recover the results of 
Ref.~\cite{Stetcu:2010xq}, 
$p\ \mbox{cot}\delta_0$ is replaced by the ERE evaluated at $p^2/m=3\omega/2$,
\begin{displaymath}
p\ \mbox{cot}\delta_0=-\frac{1}{a_0}+\frac{3}{2}\frac{r_0}{2}m\omega+.\ .\ .
\ ,
\end{displaymath}
and eq.~(\ref{eqn:s-wave ground state}) is then re-arranged in powers of
$b^{-1}$.}.
The first excited S-wave state is located at
\begin{multline}
\label{eqn:s-waveexcitedstate}
\frac{E_1^{(0)}}{\omega}=\frac{7}{2}\\
-\frac{1}{bp\ \mbox{cot}\delta_0}\left(\frac{3}{\sqrt{2\pi}}
+\frac{3(6\ \log 2-5)}{4\pi\ bp\ \mbox{cot}\delta_0}
-\frac{3\left(3\pi^2-4(11+9\ \log 2(3\ \log 2-5))\right)}{6\sqrt{2}\pi^{3/2}
\left(bp\ \mbox{cot}\delta_0\right)^2}
\ +\ .\ .\ .
\right)
\ ,
\end{multline}
where $p\ \mbox{cot}\delta_0$ is evaluated at $p^2/m=7\omega/2$.
The finite range corrections to these expressions can be introduced by
replacing $p\cot\delta_0\rightarrow p\cot\delta_0 + A\omega^2+...$ for small
$\omega$, where the
finite-range corrections depend upon the interaction, and cannot be determined
from scattering parameters alone.  It makes little sense to continue the
expansion in $\sqrt{m \omega}/(p\cot\delta)$ to higher orders
due to the appearance of the range corrections.

The same expansion can be applied to the P-waves, for which the expansion is in
terms of  $(m \omega)^{3/2}/(p^3 \text{cot}\delta_1) \ll 1$.  The lowest-lying continuum
state is located at
\begin{eqnarray}
\label{eqn:p-wavegroundstate}
\frac{E_0^{(1)}}{\omega} & = & \frac{5}{2}-\frac{3 \sqrt{\frac{2}{\pi }}}{b^3
   p^3\text{cot}\delta_1}-\frac{6 (3\log 2-4)}{\pi  \left(b^3p^3 \text{cot}\delta_1\right)^2}
\    + \ .\ .\ .
\ ,
\end{eqnarray}
with $p^3 \text{cot}\delta_1$ evaluated at $p^2/m=5\omega/2$,
the first excited state is located at 
\begin{eqnarray}
\label{eqn:p-waveexcitedstate}
\frac{E_1^{(1)}}{\omega} & = & \frac{9}{2}-\frac{\frac{15}{\sqrt{2\pi }}}{b^3
   p^3\text{cot}\delta_1}-\frac{15 (30\log 2-31)}{4\pi  \left(b^3p^3
     \text{cot}\delta_1\right)^2}
 \   + \ .\ .\ .
\ ,
\end{eqnarray}
with $p^3 \text{cot}\delta_1$ evaluated at $p^2/m=9\omega/2$, 
and the second excited state is located at
\begin{eqnarray}
\label{eqn:p-wavenextexcitedstate}
\frac{E_2^{(1)}}{\omega} & = & \frac{13}{2}
-
\frac{\frac{105}{4\sqrt{2\pi }}}{b^3 p^3\text{cot}\delta_1}
-
\frac{105 (410\log 2-389)}{128\pi  \left(b^3p^3 \text{cot}\delta_1\right)^2}
 \   + \ .\ .\ .
\ ,
\end{eqnarray}
with $p^3 \text{cot}\delta_1$ evaluated at $p^2/m=13\omega/2$.

The limit of large scattering length, $a/b\gg 1$, and small 
range, $r/b\ll 1$, the unitary limit, can also be considered.  
An expansion in powers of 
$p\cot\delta_0/(m\omega)^{1/2}$ can be performed,
and the lowest-lying S-wave state is located
(by expanding about the poles in the denominator of eq.~(\ref{eq:NNHO}) )
at
\begin{eqnarray}
{E_0^{(0)}\over\omega} & = & 
{1\over 2}\ +\ \sqrt{2\over\pi}\ {p\cot\delta_0\over\sqrt{m\omega}}\ +\ .\ .\ .\ ,
\end{eqnarray}
where $p \text{cot}\delta_0$ is evaluated at $p^2/m=\omega/2$.  This generalizes the results of Ref.~\cite{Busch:1997}.

\section{A Toy model}
\noindent
For a harmonic potential with an arbitrary value of $\omega$ 
one must rely on eq.~(\ref{eq:NNHO}) 
to extract continuum phase shifts and scattering parameters from the location
of the energy-eigenvalues, while keeping in mind that finite-range effects are
present and will move the calculated phase shift away from its true value.
To test the utility of eq.~(\ref{eq:NNHO}) in the presence 
of a finite-range interaction and to develop a ``feel'' for the size of the
finite-range corrections, we use the toy example of two particles 
interacting via a spherical well.  To make this system as `nuclear-like' 
as possible, the depth and width of the well are tuned to reproduce gross 
features of the deuteron system.  In particular, with a well depth 
$V_0=48$ MeV and radius $R_0$=1.7 fm, the single bound state has a binding  
energy $B_0$=2.22 MeV.  The scattering phase shift for this potential
is known to be
\begin{equation}
\label{eqn:wellphaseshift}
\delta_0=\text{tan}^{-1}\left[\sqrt{\frac{E_{\rm lab}}{E_{\rm lab}+2V_0}}\ 
\text{tan}\left(\sqrt{R_0\mu\left(E_{\rm lab}+2V_0\right)}\right)\right]
-\sqrt{E_{\rm lab}R_0^2V_0\mu}
\ ,
\end{equation}
where  $\mu$ is the reduced mass.

This system is placed within a harmonic potential and 
the resulting two-body spectrum at various oscillator frequencies is 
determined  numerically.
For each oscillator frequency, the spectrum is used to extract the scattering 
phase shift by virtue of eq.~(\ref{eq:NNHO}).  Because the spectrum 
is discretized, the extracted phase shifts occur at discrete points.  
By varying the oscillator frequency, the energies
at which the phase shift is determined vary thereby 
allowing 
for the energy-dependence of the phase shift to be mapped out.
\begin{figure}[!ht]
\centering
\includegraphics[width=.5\columnwidth,angle=0]{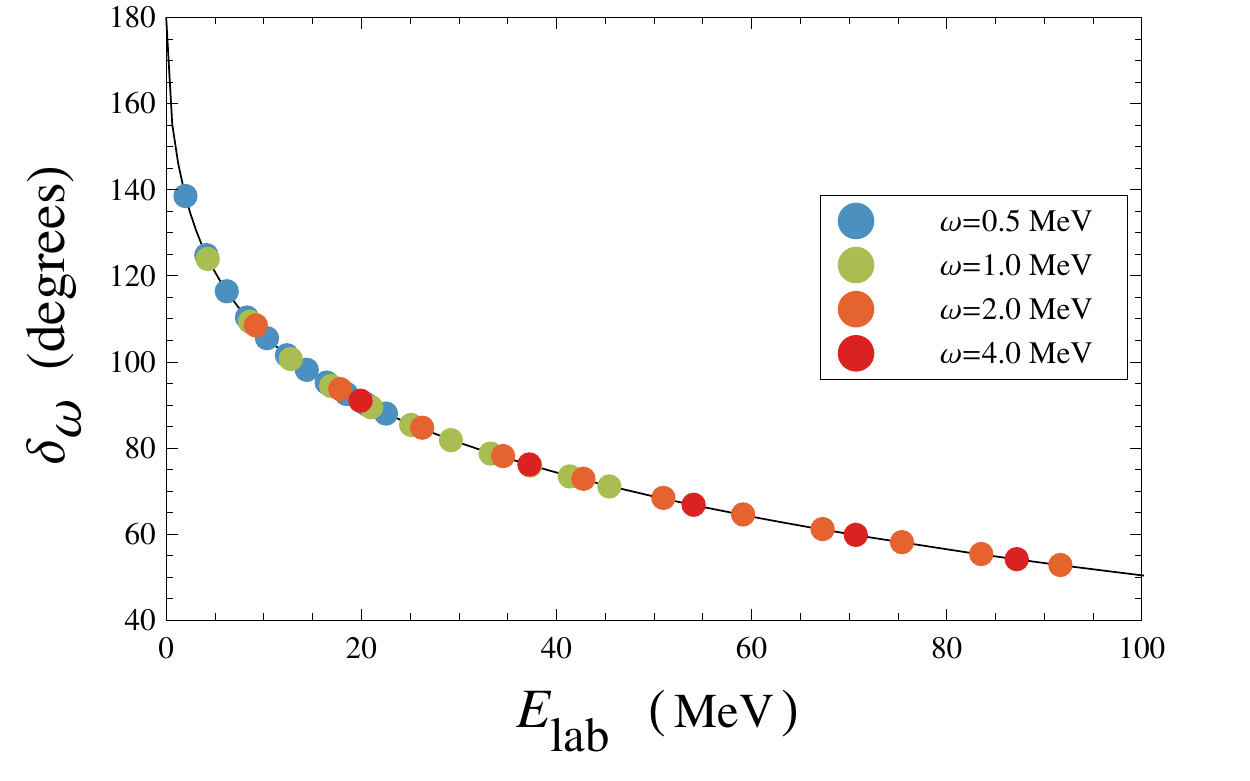}\includegraphics[width=.5\columnwidth,angle=0]{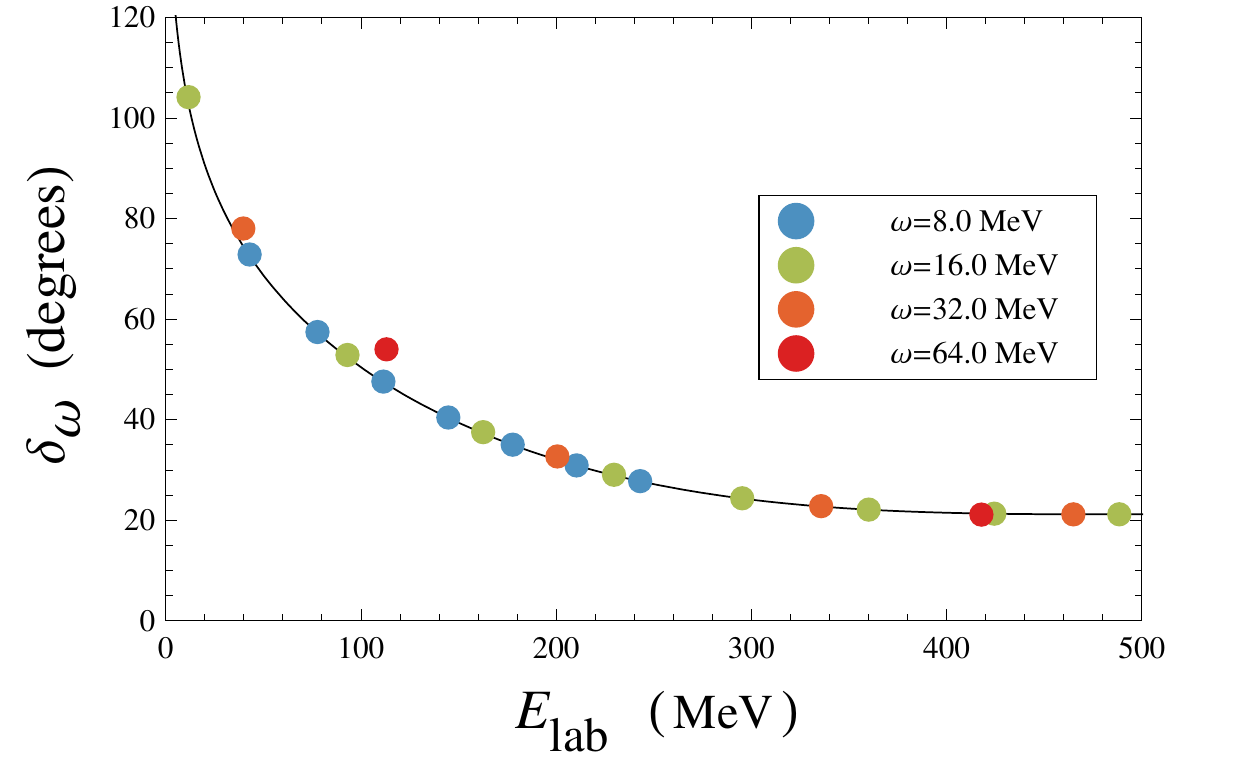}
\caption{(Color online) Extracted phase-shifts for the spherical-well toy-model 
for oscillator frequencies from  $\omega=0.5~{\rm MeV}$ to 
$\omega=4.0~{\rm MeV}$ (left panel), and  from $\omega= 8.0~{\rm MeV}$ to 
$\omega= 64.0~{\rm MeV}$ (right panel).
For each oscillator frequency,
the phase shift was determined from the 
lowest eleven energy-eigenvalues (excluding the bound state).
The exact continuum phase shift, 
given by eq.~(\protect\ref{eqn:wellphaseshift}), is the solid black curve.  
Appreciable deviations in the phase shift at larger 
oscillator frequencies are due to the finite range of the spherical well. 
\label{fig:omegawell}}
\end{figure}

\begin{figure}
\centering
\includegraphics[width=0.5\textwidth,angle=0]{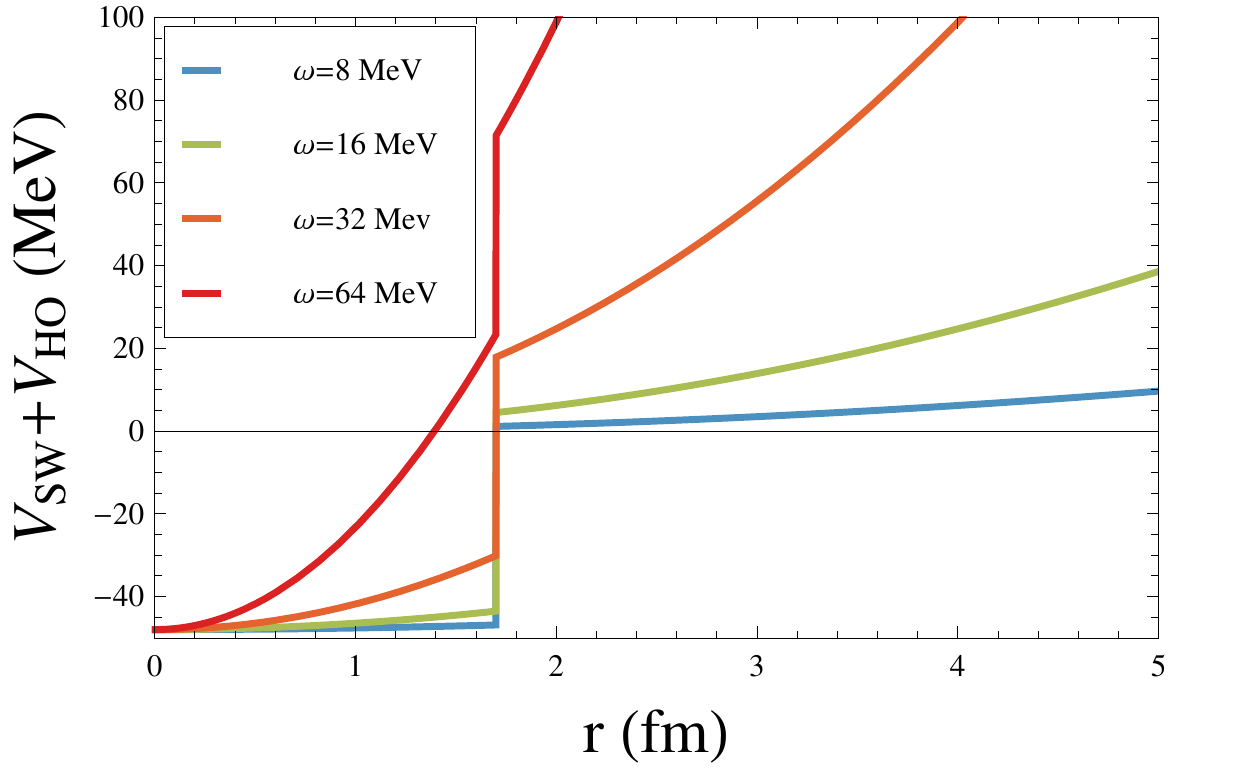}
\caption{(Color online) The potential between two particles interacting via a spherical well, 
$V_{SW}$, 
confined by a harmonic potential, $V_{HO}$,   for different 
oscillator frequencies.  As the oscillator frequency increases the distortion
of  the spherical well increases.
\label{fig:omegapluswell}}
\end{figure}

For modest-sized oscillator frequencies ($\omega < 4~{\rm MeV}$)  
the extracted phase shifts 
agree well with the exact result given in  eq.~(\ref{eqn:wellphaseshift})
(within $0.1\%$), as shown in Fig.~\ref{fig:omegawell},  
as the effects of the harmonic potential are negligible within the 
range of the spherical well.  The situation changes, 
however, for large oscillator frequencies, also shown in
Fig.~\ref{fig:omegawell}.  
In this case the exact  phase shifts  and extracted phase shifts have
appreciable differences due to the finite range of the spherical well.  
Not surprisingly, the confining nature of these potentials distorts the 
interaction of the two particles within the spherical well, 
which is demonstrated in Fig.~\ref{fig:omegapluswell}.  An interesting 
feature of these finite-range effects is that for a given oscillator 
frequency, 
the effects are largest at lower energy, and diminish as the 
energy of the system increases~\footnote{
In the high-energy limit, in which the nucleon wavelength inside
the range of the nuclear interaction is small compared to the length 
scale over
which the potential varies significantly, the LO contribution 
of the harmonic
potential to the s-wave phase shift, calculated in the WKB approximation, 
is
\begin{eqnarray}
\delta_\omega(E) 
\ -\ 
\delta_{\omega=0}(E) 
& = & 
{1\over 2\sqrt{2}}\ \mu^{3/2}\ \omega^2
\int_0^\infty\ dx\ x^2\ 
\left[\  
{1\over \sqrt{ E - V_{NN}(x)\ }}
\ -\ 
{1\over \sqrt{ E }} 
\ \right]
\ \ \ ,
\end{eqnarray}
where $\mu$ is the reduced mass of the two-nucleon system, and $V_{NN}(x)$ is
the (central) NN potential.
In the case of a toy model of NN interactions where 
a spherical well of depth $V_0$ and radius $R_0$ is used to describe the
NN potential ($V_0$ and $R_0$ are tuned to reproduce the
scattering length and effective range), the correction to the phase shift
is
\begin{eqnarray}
\delta_\omega(E) 
\ -\ 
\delta_{\omega=0}(E) 
& \rightarrow & 
{1\over 4\sqrt{2}}\ \left({\mu\over E}\right)^{3/2}\ \omega^2
\int_0^\infty\ dx\ x^2\ V_{NN}(x)
\ =\ 
{1\over 12\sqrt{2}}\ \left({\mu\over E}\right)^{3/2}\ \omega^2\ V_0\ R_0^3
\ \ \ .
\end{eqnarray}
}.  

\section{Realistic Nuclear Forces}
\noindent
It is important to determine how well this method works for 
realistic NN interactions.  There are a number of modern
NN potentials that could be used for this numerical comparison,
but for simplicity we work with the JISP16 
potential~\cite{Shirokov:2005bk,JISP16_web,Maris:2008ax}.
This $NN$ interaction 
is constructed so as to reproduce the measured
NN scattering phase shifts to high precision over a wide range of
energies, below $E_{\rm lab}\lsim 350~{\rm MeV}$,
and is known to
provide a good description of $p$ shell
nuclei~\cite{Shirokov:2005bk,Maris:2008ax} 
without an additional $3N$ interaction. 
It 
was developed using inverse scattering techniques, followed by off-shell
tuning with phase-equivalent transformations to describe selected light nuclear
properties up to $^{16}$O. 
Using this interaction, the spectrum of the two confined particles was found by diagonalizing the Hamiltonian in the relative
HO basis space for each partial wave.
The size of the HO basis was increased until the spectrum converged to a prescribed precision.
In order to access the lower energy phase shifts,  we decreased $\omega$ which consequently required an increase in the size of the basis space to obtain convergence.
This limited the range of small $\omega$ that we investigated ($\omega \ge 0.4$ MeV with a maximum basis dimension of 1800$\times$1800).
In order to improve convergence with increasing basis-space dimension,
the choice of the HO frequency for the basis space was adjusted independently of
the frequency of the external confining potential.

\begin{figure}[!ht]
\centering
\includegraphics[height=0.3\textwidth]{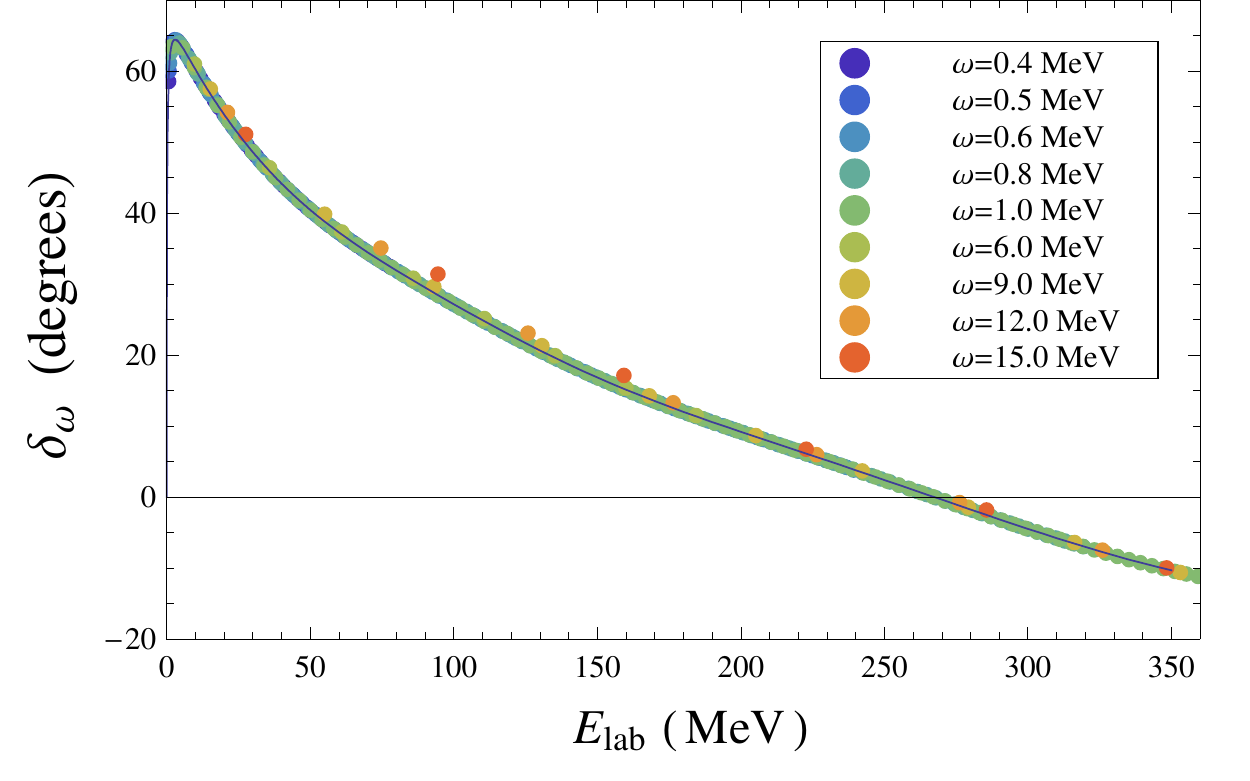}\ \ \includegraphics[height=0.3\textwidth]{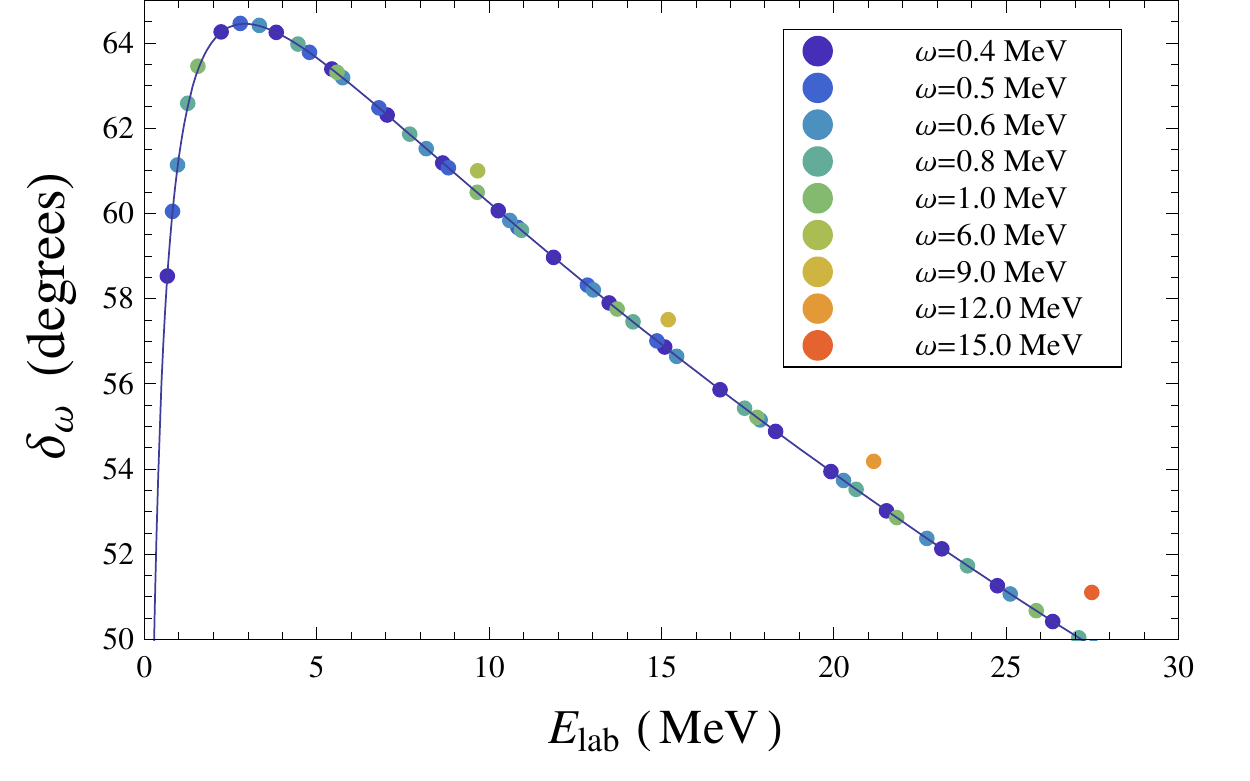}
\caption{(Color online) The phase-shift in the $^1S_0$-channel, $\delta_\omega (E_{\text{lab}})$, evaluated
  at the energy-eigenvalues determined for a range of values of $\omega$
  defining the external harmonic potential, using eq.~(\protect\ref{eq:NNHO}).
The solid curve corresponds to the phase-shift, $\delta_{\omega=0} (E_{\text{lab}})$,
determined by a direct evaluation in free-space.  
The right panel is a magnification of the left panel.
  }
\label{fig:singletdata}
\end{figure}

\begin{figure}[!ht]
\centering
\includegraphics[height=0.3\textwidth]{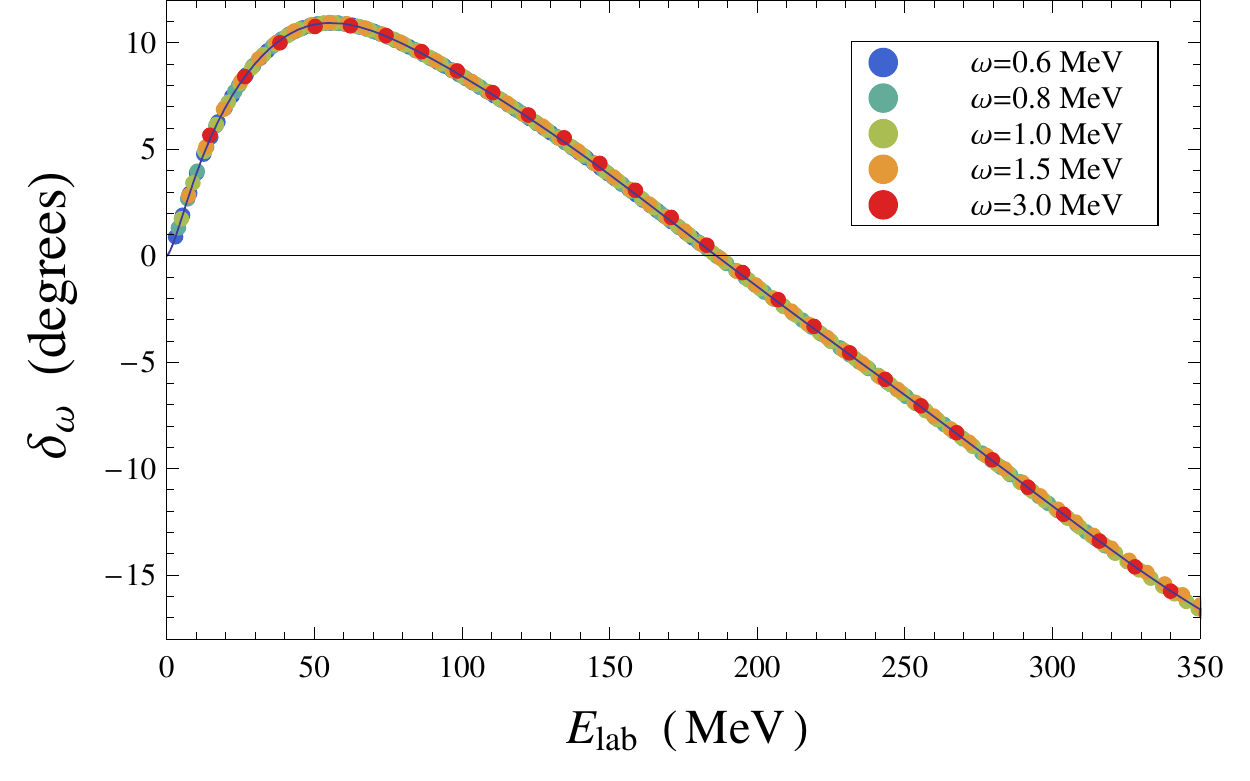}\ \ \includegraphics[height=0.3\textwidth]{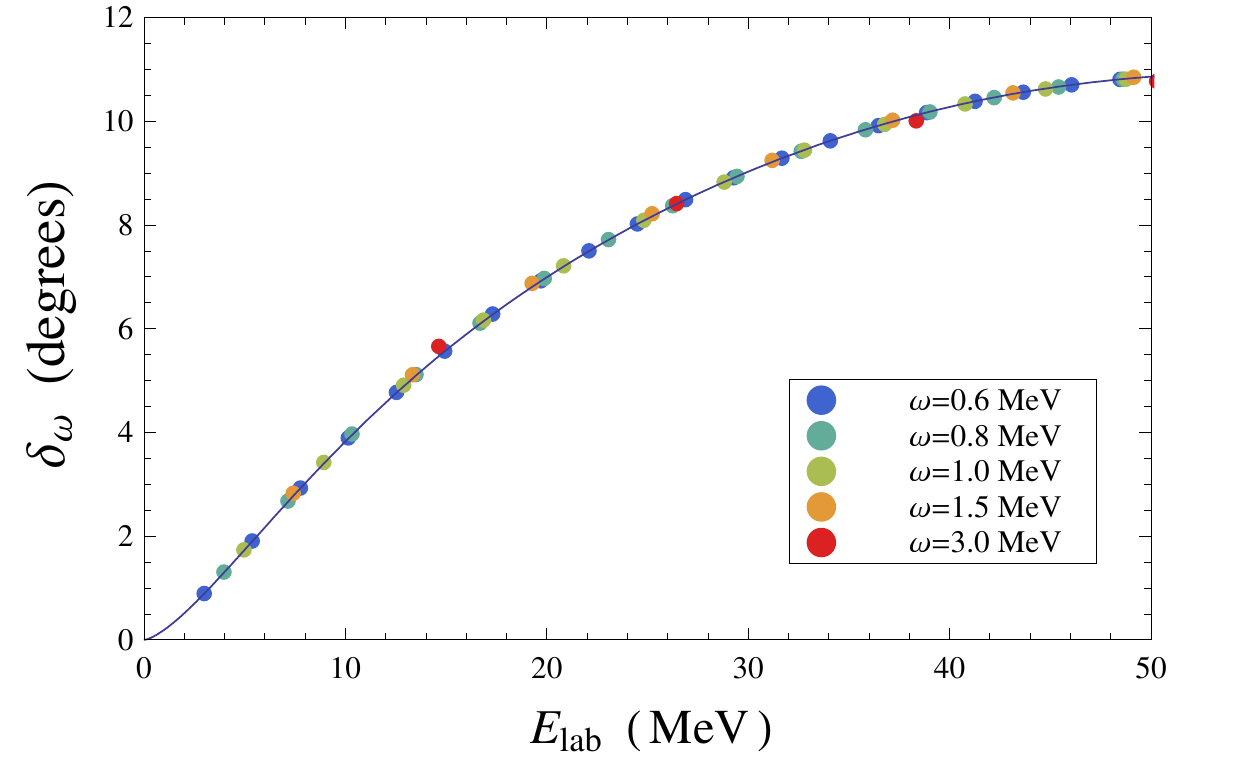}
\caption{(Color online) The phase-shift in the $^3P_0$-channel, $\delta_\omega (E_{\text{lab}})$, evaluated
  at the energy-eigenvalues determined for a range of values of $\omega$
  defining the external harmonic potential, using eq.~(\protect\ref{eq:NNHO}).
The solid curve corresponds to the phase-shift, $\delta_{\omega=0} (E_{\text{lab}})$,
determined by a direct evaluation in free-space.  
The right panel is a magnification of the left panel.
  }
\label{fig:tripletPdata}
\end{figure}

\begin{figure}[!ht]
\centering
\includegraphics[height=0.3\textwidth]{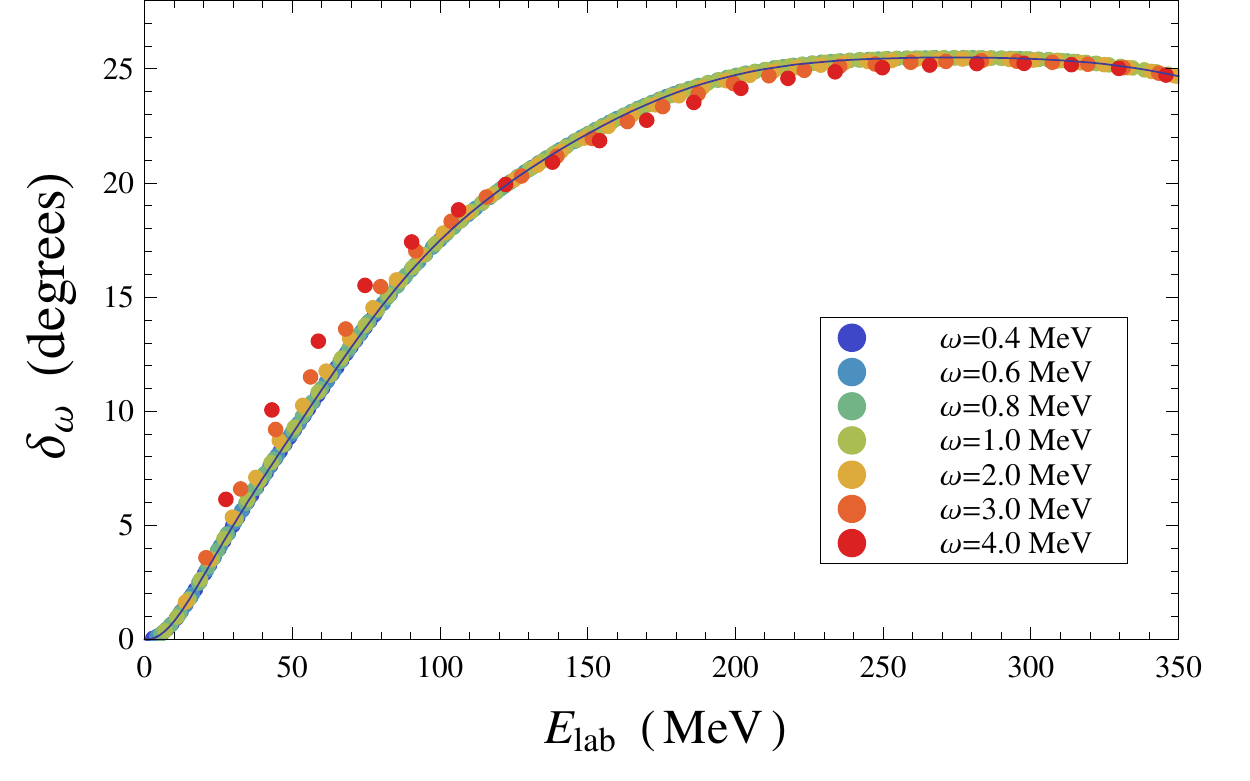}\ \ \includegraphics[height=0.3\textwidth]{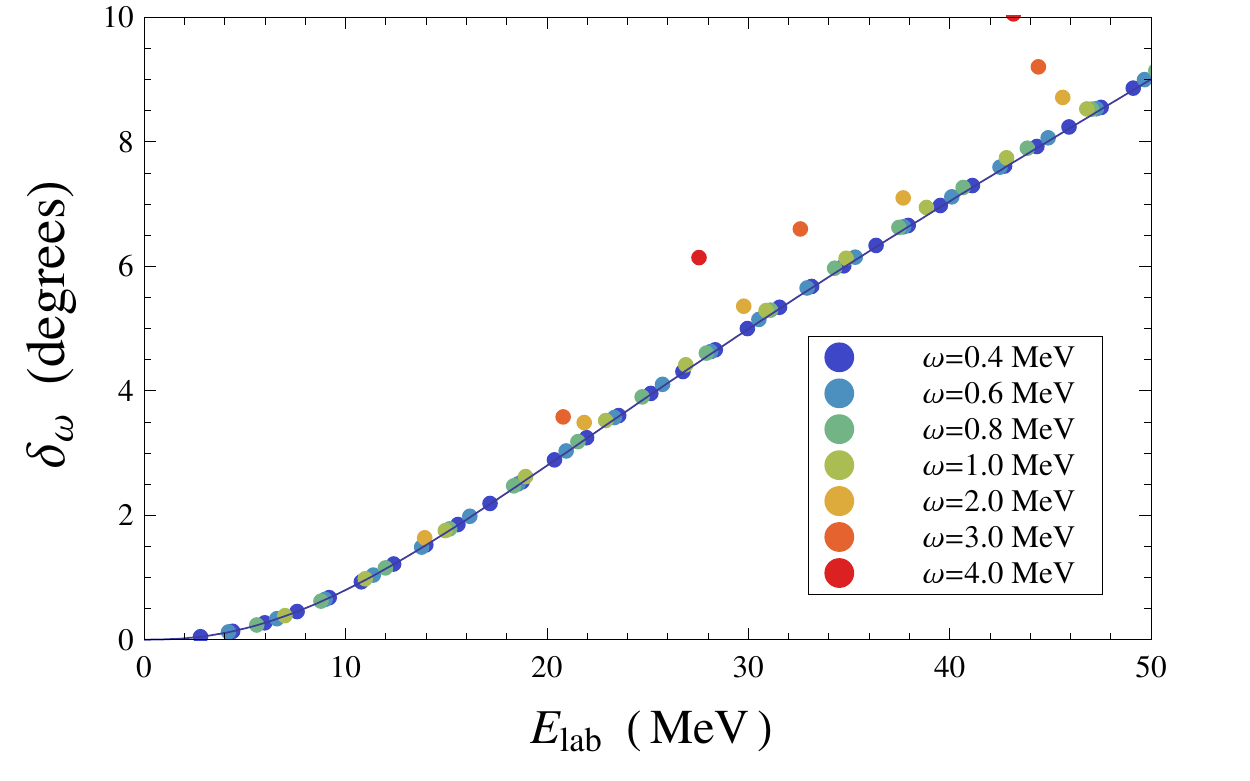}
\caption{(Color online) The phase-shift in the $^3D_2$-channel, $\delta_\omega (E_{\text{lab}})$, evaluated
  at the energy-eigenvalues determined for a range of values of $\omega$
  defining the external harmonic potential, using eq.~(\protect\ref{eq:NNHO}).
The solid curve corresponds to the phase-shift, $\delta_{\omega=0} (E_{\text{lab}})$,
determined by a direct evaluation in free-space.  
The right panel is a magnification of the left panel.
  }
\label{fig:tripletDdata}
\end{figure}

\begin{figure}[!ht]
\centering
\includegraphics[height=0.3\textwidth]{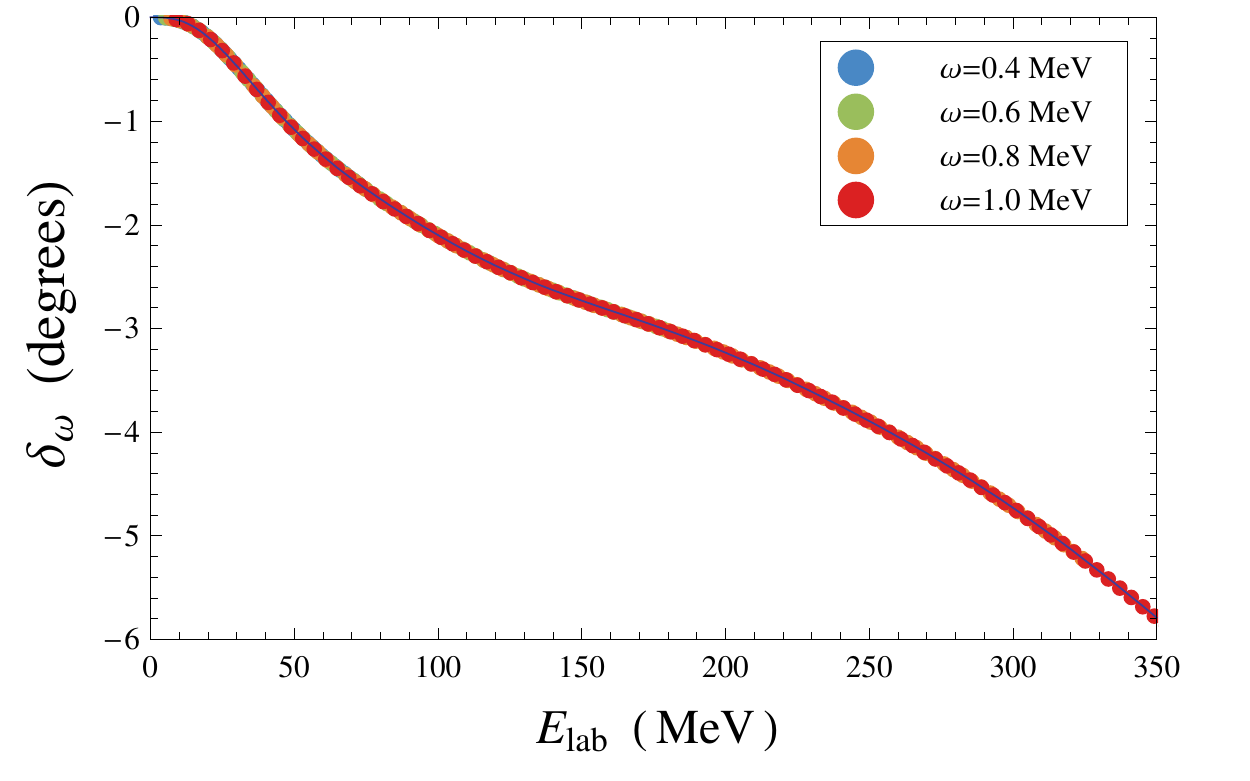}\ \ \includegraphics[height=0.3\textwidth]{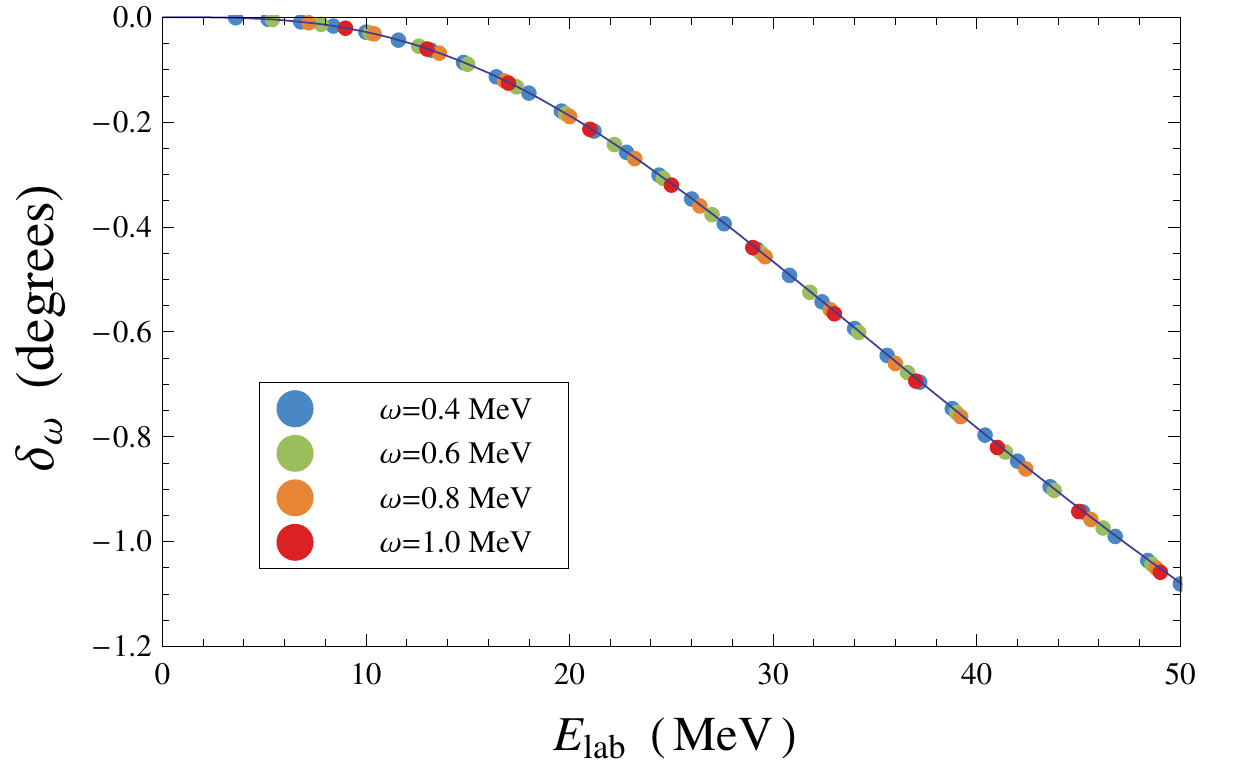}
\caption{(Color online) The phase-shift in the $^1F_3$-channel, $\delta_\omega (E_{\text{lab}})$, evaluated
  at the energy-eigenvalues determined for a range of values of $\omega$
  defining the external harmonic potential, using eq.~(\protect\ref{eq:NNHO}).
The solid curve corresponds to the phase-shift, $\delta_{\omega=0} (E_{\text{lab}})$,
determined by a direct evaluation in free-space.  In the left panel this solid curve is obscured by the points calculated using eq.~(\protect\ref{eq:NNHO}).
The right panel is a magnification of the left panel.
  }
\label{fig:singletFdata}
\end{figure}

In Figs.~\ref{fig:singletdata} to \ref{fig:singletFdata} we show the application
of  eq.~(\ref{eq:NNHO})  to four different partial waves in the $NN$ system: 
$^1S_0$ ($l=0$), $^3P_0$ ($l=1$), $^3D_2$ ($l=2$) and $^1F_3$ ($l=3$). 
The extracted phase-shifts were obtained from the low-lying 
spectrum of the  $NN$ system in harmonic potentials with a range of frequencies
(the points in each figure).
For comparison, the phase-shifts calculated by solving the Schr$\ddot{\text{o}}$dinger equation
in the absence of the harmonic potential are shown as the solid curves in each figure.

%

\subsection{Numerical Analysis}
\label{sect:numericalanalysis}
\noindent
The harmonic potential modifies the interactions between the two nucleons 
at all distance scales, and as such, there are modifications to the
potential between the nucleons over the range of the nuclear forces, 
leading to short-distance corrections to the relation between $p\cot\delta$
and the energy-eigenvalues given by eq.~(\ref{eq:NNHO}).  
The energy-eigenvalues are calculated in 
a given energy-interval for a range of
values of $\omega$ in order to extrapolate the phase-shift 
$\delta_\omega (E_{\text{lab}})$, to the $\omega=0$ limit, $\delta_{\omega=0} (E_{\rm lab})$, and
hence eliminate the modifications to the nuclear force due to the 
harmonic potential.  This procedure is not as straightforward as it naively
appears due to the fact that for each value of $\omega$, the
energy-eigenvalues (generally)
have different values, and an
interpolation of $\delta_\omega (E_{\rm lab})$
within the energy-interval is required 
for each $\omega$ in order to extrapolate 
to $\delta_{\omega=0} (E_{\rm lab})$ at any given energy~\footnote{For a given energy, a
  range of values of $\omega$ could be iteratively tuned to produce the same
  energy-eigenvalue.}.
For the sake of demonstration, we focus on the phase-shift in the 
$^1S_0$ channel, but the methodology can be applied in all channels.

The energy-eigenvalues of two nucleons interacting in the $^1S_0$ in a
harmonic potential were calculated
for a range of values of $\omega$ from
$\omega=0.4$ MeV to $\omega=15.0$ MeV.  For each eigenvalue, the
scattering phase-shift $\delta_\omega (E_{\rm lab})$ was calculated using the zero-range
relation in eq.~(\ref{eq:NNHO}), the results of which are shown in 
Fig.~\ref{fig:singletdata}.
The ``exact'' phase-shift $\delta_{\omega=0} (E_{\rm lab})$
that is determined by solving the Schr$\ddot{\text{o}}$dinger equation for the phase-shift in the absence of
the external harmonic potential is shown in Fig.~\ref{fig:singletdata} as the
solid curve.
For $\omega\ \lsim 1.0$ MeV, 
and for the energy-eigenvalues shown in Table~\ref{tab:1s0}, the
phase-shift calculated from the zero-range relation in eq.~(\ref{eq:NNHO})
is very close to the actual phase-shift.  For larger values of $\omega$ there
are noticeable deviations from the exact result, but these deviations 
are found to become
smaller as the energy increases. 
\begin{table}[!ht]
  \caption{\label{tab:1s0}{The lowest eight energy-eigenvalues in the center-of-mass frame and their associated phase-shifts found
      from eq.(\protect\ref{eq:NNHO}) in the $^1S_0$-channel for $\omega \le 1$ MeV.   }}
  \begin{ruledtabular}
    \begin{tabular}{ccccccc}
        & $\omega=0.4~{\rm MeV}$   & $\omega=0.5~{\rm MeV}$ & $\omega=0.6~{\rm
          MeV}$  
& $\omega=0.8~{\rm MeV}$  & $\omega=0.9~{\rm MeV}$  & $\omega=1.0~{\rm MeV}$  \\
      \hline
$E_1$ & 0.66642 & 0.81618 & 0.96488& 1.2610&1.40898&1.55711\\
$\delta_\omega (E_1)$ & 58.5279 & 60.0449& 61.1382& 62.5816&63.0684&63.4511\\
\hline
$E_2$ & 2.22732 & 2.78198& 3.33893& 4.4597&5.02344&5.58933\\
$\delta_\omega (E_2)$ & 64.2586 & 64.45411& 64.4124& 63.9721&63.6576&63.3088\\
\hline
$E_3$ & 3.82836 & 4.7907& 5.75672& 7.69914& 8.67518&9.65424\\
$\delta_\omega (E_3)$ & 64.2495 & 63.7768& 63.1847& 61.8606& 61.1775&60.4948\\
\hline
$E_4$ & 5.4363 & 6.80548& 8.17932& 10.9398& 12.326&13.7158\\
$\delta_\omega (E_4)$ & 63.3856 & 62.4755& 61.5194& 59.6034& 58.6685&57.754\\
\hline
$E_5$ & 7.04598 & 8.82126& 10.602&14.1783& 15.9733&17.7726\\
$\delta_\omega (E_5)$ & 62.3082 & 61.0673& 59.832&57.4512& 56.3152&55.214\\
\hline
$E_6$ & 8.65604 & 10.8369& 13.0238&17.4144& 19.6173&21.825\\
$\delta_\omega (E_6)$ &  61.1822 & 59.6694& 58.2013&55.4272& 54.1198&52.8615\\
\hline
$E_7$ & 10.266 & 12.852& 15.4446&20.6482& 23.2584&25.8736\\
$\delta_\omega (E_7)$ &  60.0611& 58.3139& 56.6418&53.5231& 52.0696&50.6769\\
\hline
$E_8$ & 11.8758 & 14.8665 & 17.8645&23.8802& 26.897&29.9188\\
$\delta_\omega (E_8)$ & 58.9638 & 57.0076& 55.1527& 51.7288& 50.1421&48.6454\\
\hline
    \end{tabular}
  \end{ruledtabular}
\end{table}

The energy-dependent interpolation of the phase-shift for a given $\omega$ 
that is required in
order to extrapolate $\delta_\omega (E_{\rm lab})$ to $\delta_{\omega=0} (E_{\rm lab})$ is the most
problematic part of this numerical analysis.
In the range of energies for  which the ERE is formally
convergent ($|{\bf p}|\le m_\pi/2$ resulting from the location of the t-channel
cut in the one-pion exchange amplitude)
an expansion of $p\cot\delta$ in powers of the energy
reproduces the scattering amplitude.  
In contrast, for energies outside
of this range, but below the inelastic threshold 
(for which the relation between $p\cot\delta$ and the
energy-eigenvalues in the harmonic potential in  eq.~(\ref{eq:NNHO}) remains valid) 
such a power-series does not describe the
amplitude.  
As such, without directly solving for the amplitude, one does not
know the form to use for the interpolation in relative energy $E$ 
beyond $|{\bf p}| = m_\pi/2$.  
We do not attempt to resolve this issue, and restrict ourselves to the
energy range for which the ERE formally converges \footnote{Within this range, this part of our analysis is formally equivalent to the pionless EFT description given in Refs.~\cite{bira:exascale,Stetcu:workshop}.}.
\begin{figure}[!ht]
\centering
\includegraphics[height=0.3\textwidth]{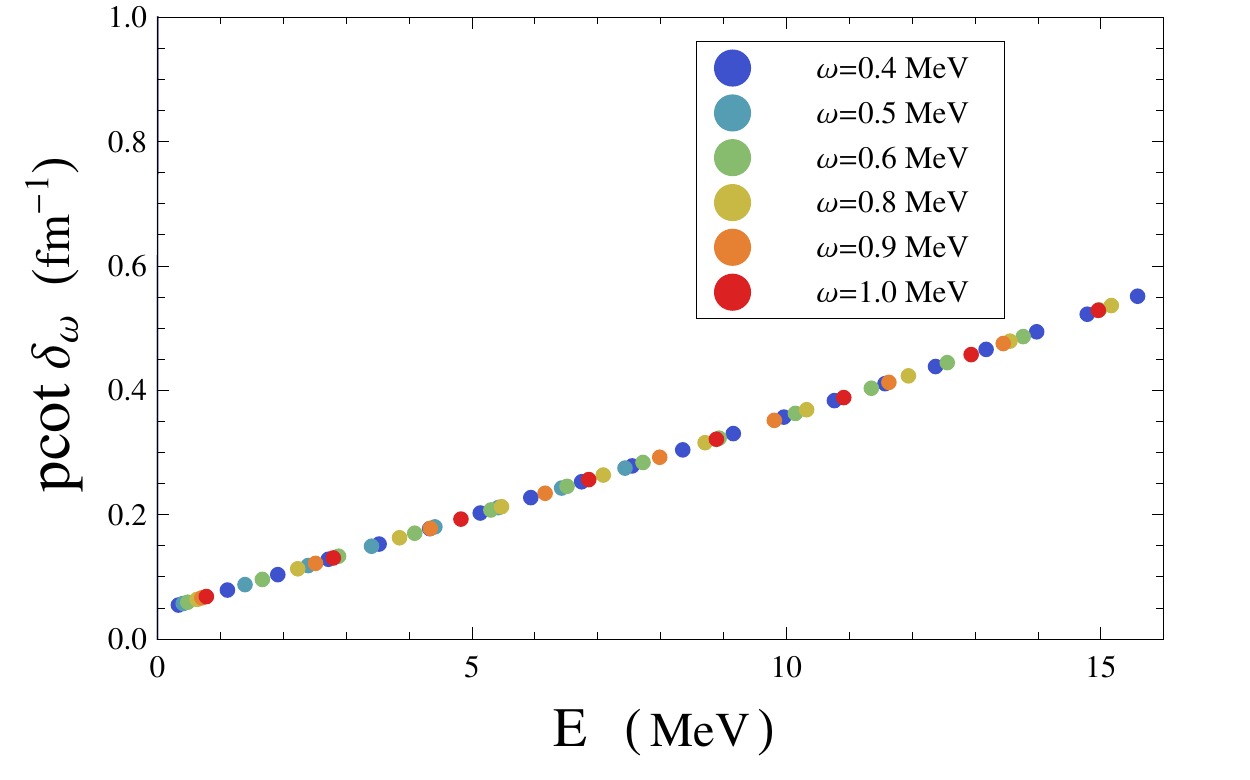}
\caption{(Color online) $p\cot\delta$ as a function of the energy in the center-of-mass
 frame in the $^1S_0$-channel
extracted from the energy-eigenvalues of the two-nucleon system
in harmonic potentials for a range of  oscillator frequencies $\omega$.
  }
\label{fig:singletPCOTDdata}
\end{figure}
In Fig.~\ref{fig:singletPCOTDdata} we show the extracted values of
$p\cot\delta$ as a function of relative energy $E$ for $\omega\le 1$ MeV.
For each $\omega\le 1$ MeV a fourth-order polynomial in $E$ is fit 
to the values of $p\cot\delta$
shown in Fig.~\ref{fig:singletPCOTDdata} (the order was chosen to minimize the
$\chi^2/{\rm dof}$ of the fit and to achieve a stable fit under the change of
order~\footnote{A full systematic study of uncertainties would include the
  variation of the fit with the polynomial order.}  ).
With the interpolating functions, it is then possible to choose a particular
value of $E$ and extrapolate $p\cot \delta_\omega (E)$ to $p\cot
\delta_{\omega=0} (E)$, from which the phase-shift $\delta_{\omega=0} (E)$
can be recovered.
\begin{figure}[!ht]
\centering
\includegraphics[height=0.3\textwidth]{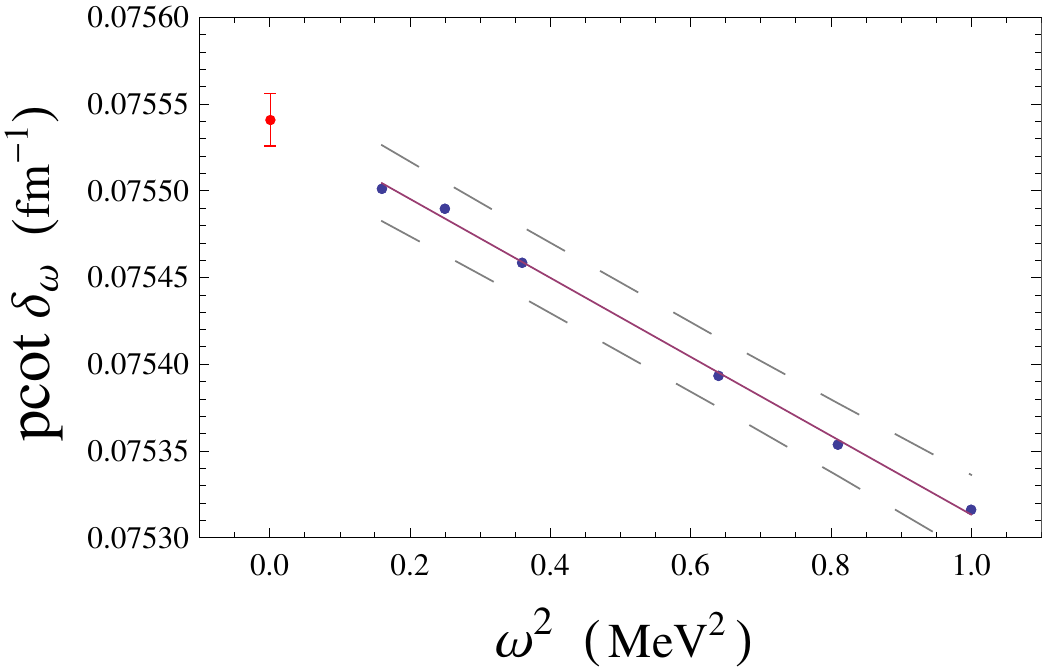}\ \ \includegraphics[height=0.3\textwidth]{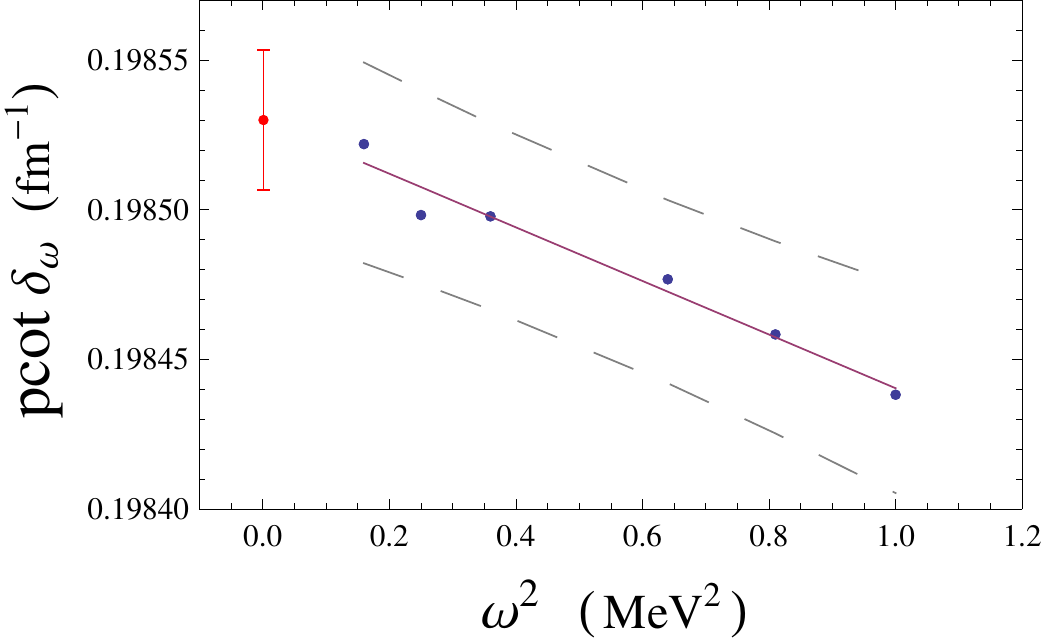}
\caption{(Color online) The extrapolation of $p\cot \delta_\omega (E)$ to  
$p\cot\delta_{\omega=0} (E)$ 
in the $^1S_0$-channel at $E=1~{\rm MeV}$ (left panel) and  $E=5~{\rm MeV}$ (right panel).    
The solid lines correspond to the best fits of the form $p\cot\delta_\omega = A + B\  \omega^2$, 
and the dashed lines correspond to the $99\%$ confidence intervals.
The red points with their associated 1-sigma uncertainties
correspond to $\delta_{\omega=0} (E)$.  }
\label{fig:omegaextrap}
\end{figure}
The $\omega$-extrapolations at $E=1~{\rm MeV}$  and 
$E=5~{\rm MeV}$ are shown in Fig.~\ref{fig:omegaextrap}.  
A fit function of the form $p\cot\delta_\omega = A + B\  \omega^2$ is 
used to extrapolate to $\omega=0$, as also shown in Fig.~\ref{fig:omegaextrap}.  
The small observed scatter of the points about the best fit function 
is attributed to the form of the interpolation in $E$ (and the increasing
separation between energy-eigenvalues with increasing $\omega$), and not the
finite model-space as the energy-eigenvalues have converged to six significant digits  which we establish by increasing $N_{MS}$
sufficiently.
An important point to note is that the results of the calculations at the
smallest few values of  $\omega$ are all within  $\sim 0.1\%$ of the
extrapolated values. Therefore, to determine the phase-shift at this level of precision,
no extrapolation in $\omega^2$ is required.
\begin{figure}[!ht]
\centering
\includegraphics[height=0.6\textwidth]{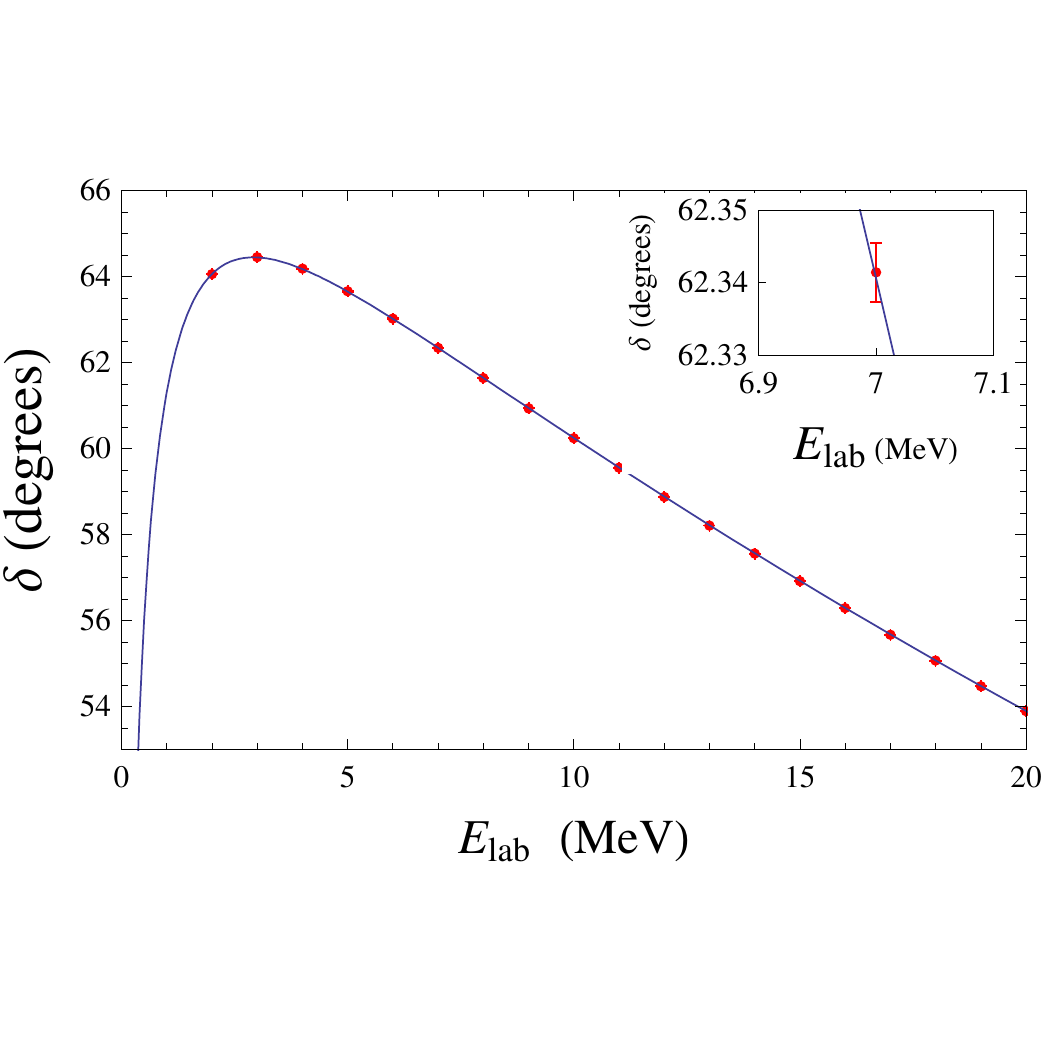}
\caption{(Color online) The $\omega$-extrapolated phase-shift $\delta_{\omega=0} (E_{\rm lab})$ as a
  function of $E_{\rm lab}$ in the $^1S_0$-channel.  
The points and their uncertainty are determined, at uniform intervals in $E_{\rm lab}$, 
by the interpolations and
extrapolations described in the text.
The solid curve corresponds to the ``exact'' phase-shift.
The inset is a magnification around $E_{\rm lab}=7~{\rm MeV}$ that
  shows the precision of the calculation.
  }
\label{fig:deltaextrapSING}
\end{figure}
The extrapolated phase-shift $\delta_{\omega=0} (E_{\rm lab})$ in the $^1S_0$-channel
is shown in Fig.~\ref{fig:deltaextrapSING},
and is found to agree with the exact phase-shift (the solid curve)
within the uncertainties of the
calculation.
The points with uncertainties correspond to the phase-shift derived from the
energy-eigenvalues extrapolated to $\omega=0$
evaluated at regular intervals in $E$. 
Uncertainties in the extrapolated phase-shifts, which are at the $\sim
10^{-4}$-level,  can, in principle, be reduced further
by calculating at even smaller values of $\omega$.

\begin{figure}[!ht]
\centering
\includegraphics[height=0.32\textwidth]{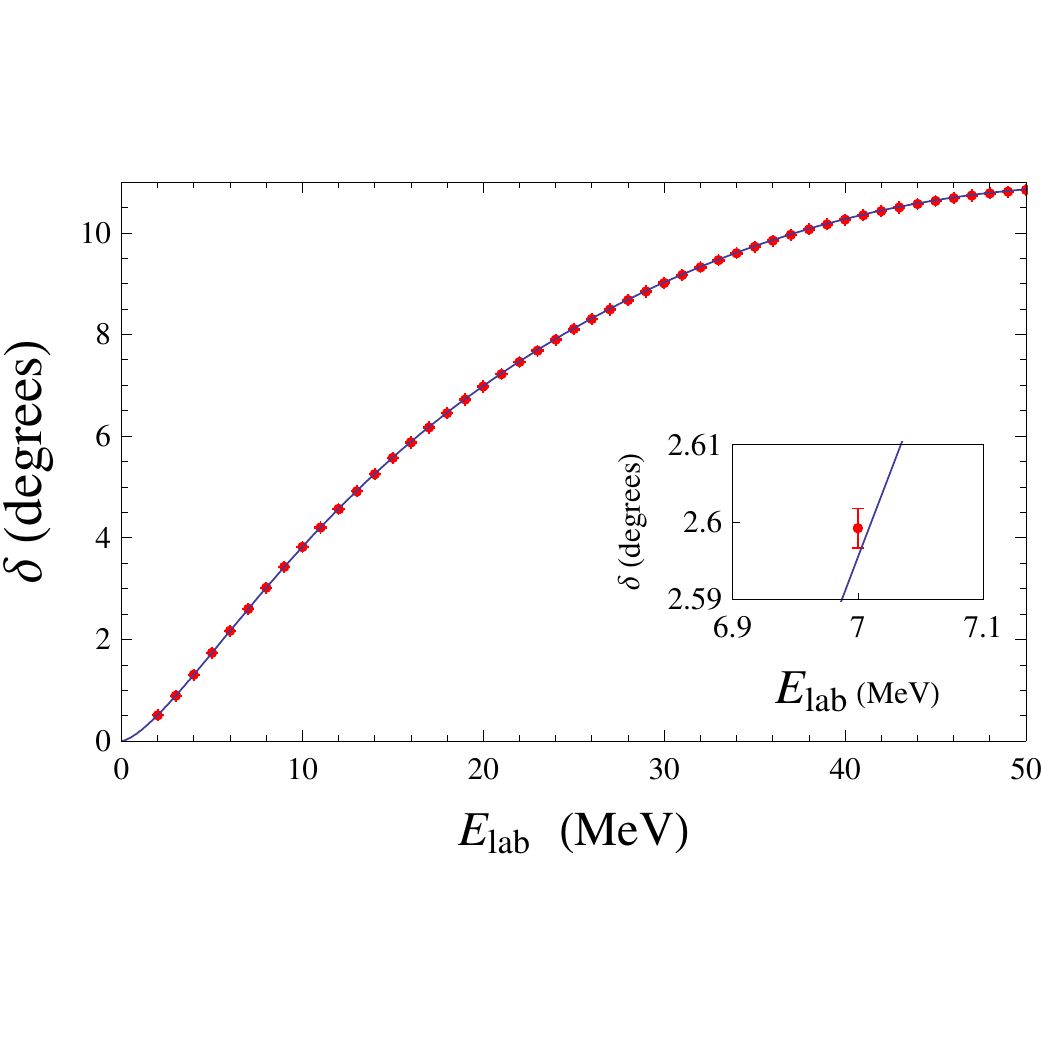}\
\includegraphics[height=0.32\textwidth]{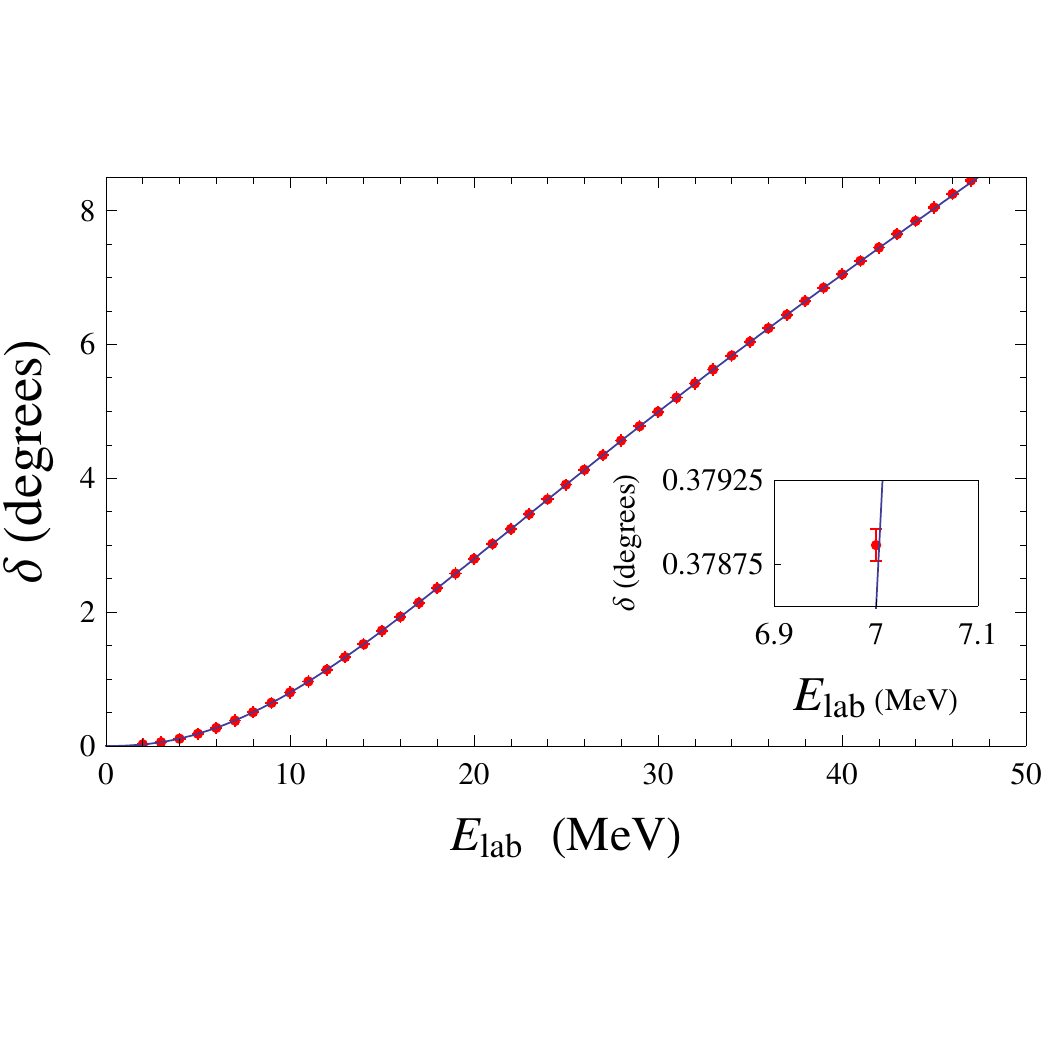}\ \includegraphics[height=0.32\textwidth]{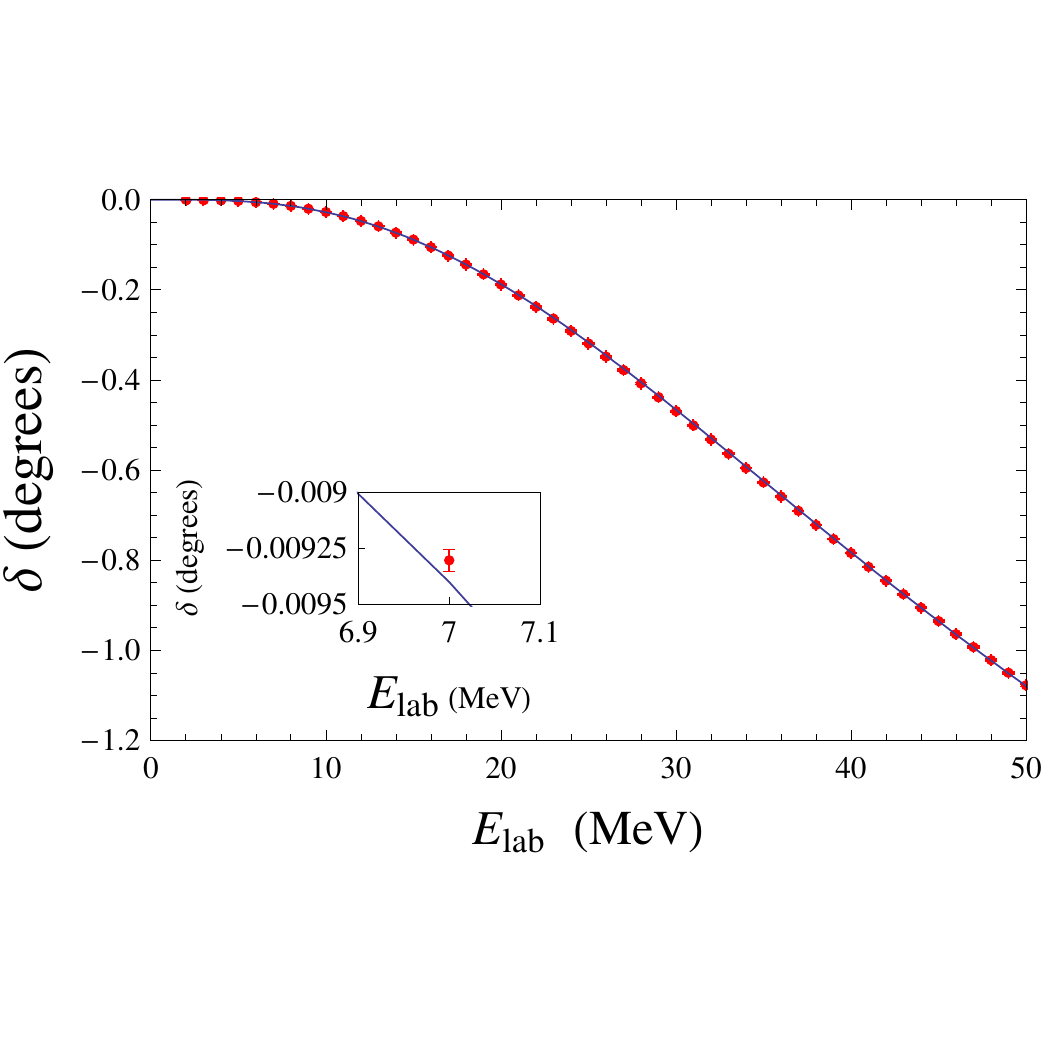}
\caption{(Color online) The $\omega$-extrapolated phase-shift $\delta_{\omega=0} (E_{\rm lab})$ as a
  function of $E_{\rm lab}$ in the 
$^3P_0$-channel (left panel),  the $^3D_2$-channel (center panel) and the 
$^1F_3$-channel (right panel).  
The insets are a magnification around $E_{\rm lab}=7~{\rm MeV}$ that
  shows the precision of the calculations.
  }
\label{fig:deltaextrapHigher}
\end{figure}

We have numerically explored some of the higher partial-waves.
The methodology in the higher partial-waves is the same as in the
$^1S_0$-channel.
The harmonic potential modifications to the nuclear force are seen to increase
with increasing partial-wave.  This behavior is expected due to the fact that
the centripetal barrier, and the associated $r^l$ behavior of the wavefunction
near the origin, forces the wavefunction to larger values of $r$ (but within the range of the nuclear force) and hence to larger values of the harmonic potential.
Calculations at smaller values of $\omega$ than employed for the S-wave case
are required in order to achieve the same  level of precision, consistent with the conclusions of Ref.~\cite{Suzuki:2009}.
The extracted values of the phase-shift in the $^3P_0$, $^3D_2$ and $^1F_3$
channels extrapolated to $\omega=0$
are shown in
Fig.~\ref{fig:deltaextrapHigher}.  In all channels, the extrapolated
phase-shifts are  found to agree with the ``exact''
phase-shift within the uncertainties of the calculation.

\begin{figure}[!ht]
\centering
\includegraphics[height=0.32\textwidth]{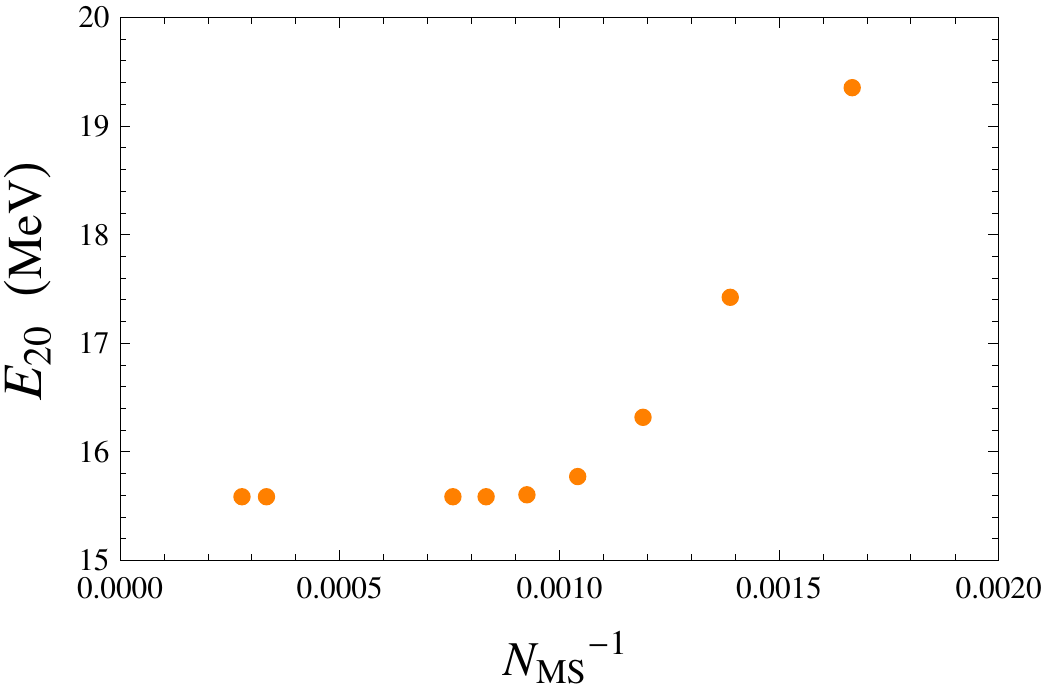}
\caption{(Color online) The $20^{\rm th}$ energy-eigenvalue in the center-of-mass frame for the $^1S_0$-channel with
  $\omega=0.4~{\rm MeV}$ as a function of the inverse cut-off of the
  model-space,
$1/N_{\rm MS}$. 
  }
\label{fig:Nconverge}
\end{figure}

Determining the energy levels of two nucleons in a harmonic
potential involves calculating the matrix elements of the full
Hamiltonian, including the harmonic 
potential, in
a large model-space, with a cutoff on relative excitation energies denoted by $\omega N_{\rm MS}$.
In the limit that $N_{\rm MS}\rightarrow\infty$ the energy-eigenvalues found
by diagonalizing the $N_{\rm MS}\times N_{\rm MS}$ Hamiltonian will
coincide with the exact energy-eigenvalues.   
For a finite-dimensional space,
the energy-eigenvalues deviate from their infinite model-space values 
as shown, for
instance,  in Fig.~\ref{fig:Nconverge},  
making the quantification
of the 
convergence of eigenvalues with respect to $N_{\rm MS}$ highly non-trivial.  
We do not attempt to resolve this issue here, and all of the 
energy-eigenvalues we
have used in this work have converged to at least six significant digits.  

In this work we have only analyzed uncoupled channels for simplicity.  In general, due to the spin of the
nucleon, and the fact that two nucleons can have $S=1$, many two-nucleon states
with total angular momentum $J$ are a linear combination of two orbital angular
momentum states, such as the $^3S_1$-$^3D_1$ coupled-channel which
contains the deuteron.
The zero-range relation between the energy-eigenvalues in the harmonic potential
and the scattering phase shift, given in eq.~(\ref{eq:NNHO}), will be modified to
be a relation involving the three scattering parameters that describe a
two-component coupled-channels system, e.g.  $\delta_0$, $\delta_2$
and $\epsilon_1$, and not just a simple relation between one phase shift and
the energy-eigenvalues.  
Such relations remain to be constructed for two nucleons in
a harmonic potential.
As the $^3S_1$-$^3D_1$ coupled-channels system contains the deuteron as a
bound-state, we can explore the behavior of the lowest energy-eigenvalue as a
function of $\omega$.  
The perturbative corrections to the location of such
bound states due to the presence of the harmonic potential are given in 
eq.~(\ref{eqn:s-waveboundstate}), where the LO deviations
scale as $\sim\omega^2$.
The binding energies are found to be 
$E_0=-2.2209~{\rm MeV}$,  $-2.2163~{\rm MeV}$,
$-2.2098~{\rm MeV}$,  and $-2.2017~{\rm MeV}$ in harmonic potentials with 
$\omega=0.2~{\rm MeV}$, $0.3~{\rm MeV}$, $0.4~{\rm MeV}$ and $0.5~{\rm MeV}$, respectively.
\begin{figure}
\centering
\includegraphics[height=0.4\textwidth]{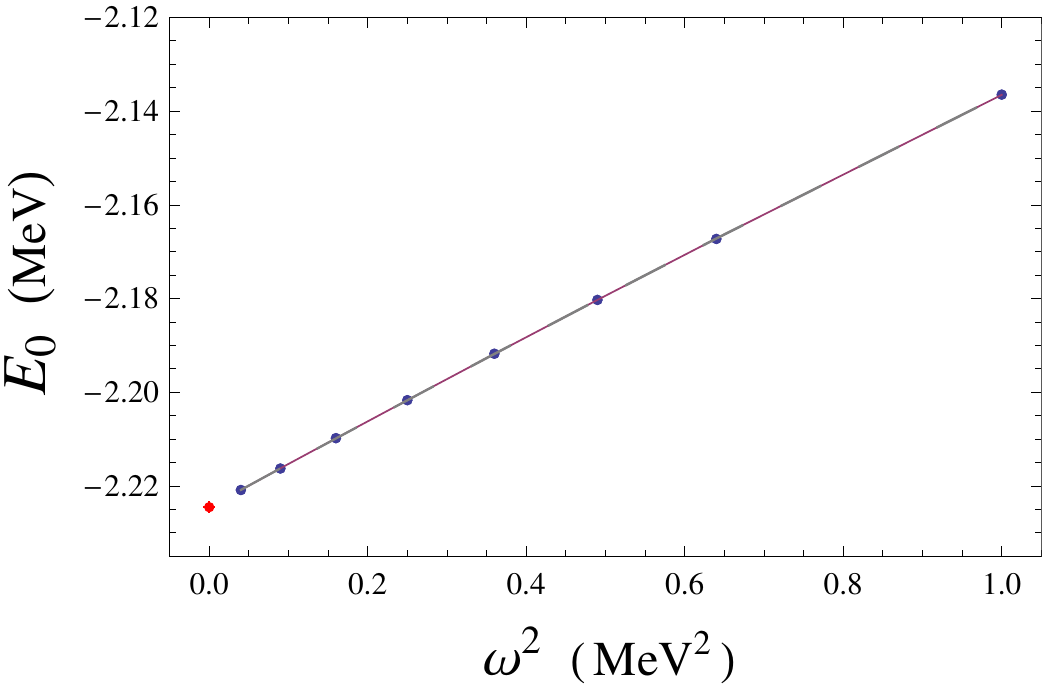}
\caption{(Color online) The deuteron binding energy as a function of $\omega^2$.  The solid
  line corresponds to the best fit of the form 
$E_0 = E_0^{(\omega=0)} + C\ \omega^2 + D\ \omega^4 + F\ \omega^6$, 
and the dashed lines (practically indistinguishable from the solid line) denote the $68\%$ confidence interval.  The red point corresponds to the ground state energy obtained by extrapolating to $\omega=0$.  The uncertainty is within the size of the red point. }
\label{fig:bindingextrap}
\end{figure}
The extrapolation of these values to $\omega=0$ is shown in
Fig.~\ref{fig:bindingextrap}.
The results are fit well by a polynomial of the form
$E_0 = E_0^{(\omega=0)} + C\ \omega^2 + D\ \omega^4 + F\ \omega^6$, 
where $C,D,F$ and $E_0^{(\omega=0)}$ are  fit variables, 
for the range of $\omega$ for which the calculations have been performed.
The deuteron binding energy extracted from the extrapolation to $\omega=0$ is 
$E_0^{(\omega=0)}= -2.22466(4)~{\rm MeV}$ (which is to be compared with the
input value of $-B_0= -2.224575~{\rm MeV}$). 
The coefficient of the $\omega^2$ term is $C_{\rm fit}=0.0939(4)$, which
is consistent with the value expected in the zero-range limit of $C_{\rm ZR}=0.0944$
from eq.~(\ref{eq:BEshift}).
One expects both the LO short-range
contributions from  $\omega \ne 0$ and the small D-state admixture due to the tensor force to also depend
upon $\omega^2$, and to modify the value of $C$ away from $C_{\rm ZR}$, but it
is clear from this work that such deviations are small.

\begin{figure}
\centering
\includegraphics[width=0.4\textwidth]{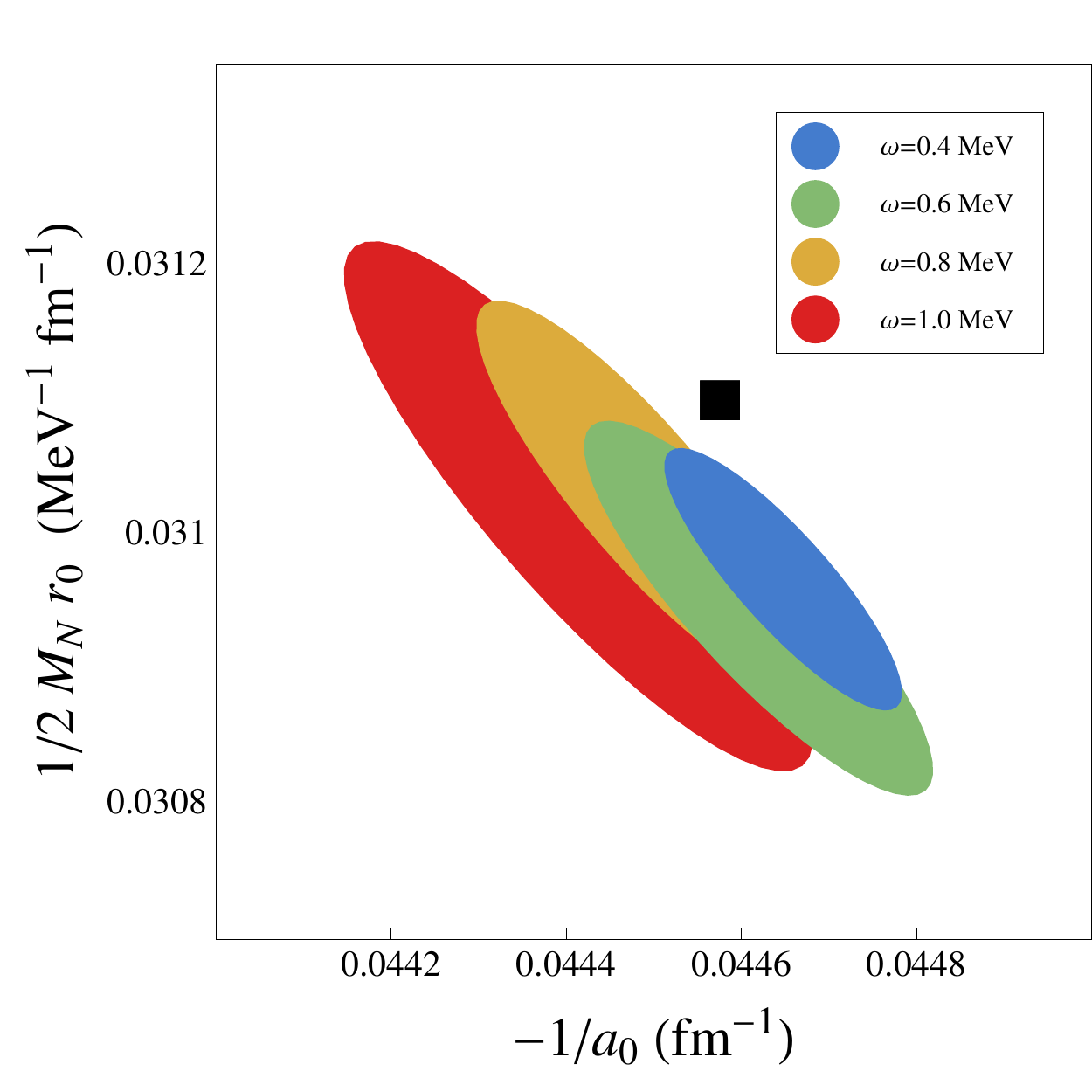}\
\ \ \ \ \ \ \ \ \ \includegraphics[width=.4\textwidth]{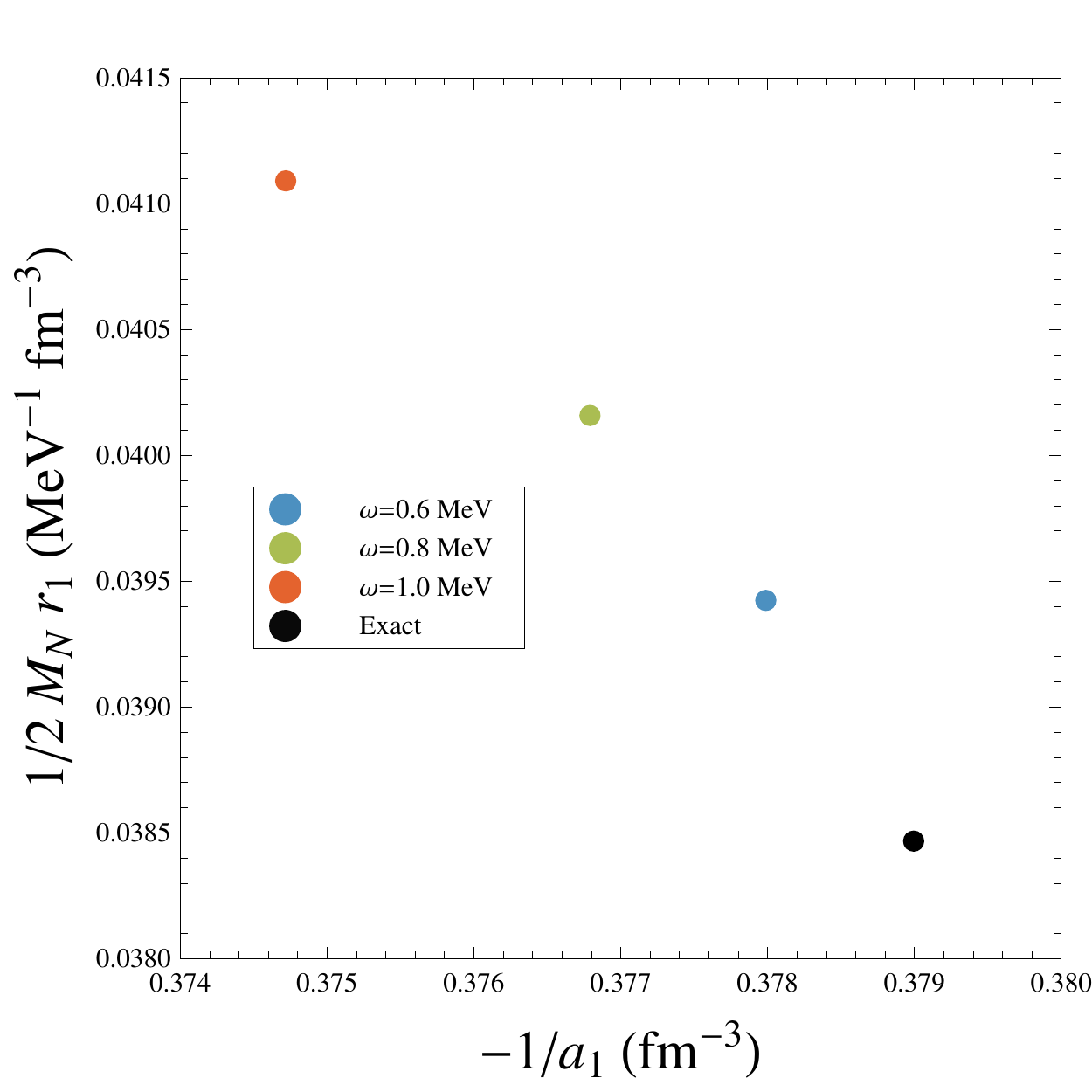}
\caption{(Color online) The extracted $^1S_0$ (left panel) and $^3P_0$ (right panel) 
scattering lengths and effective ranges for different harmonic potentials.  
The exact values  are denoted by the black points.  
The ellipses on the left panel denote the $68\%$ confidence intervals and 
were determined by fitting the ERE to the interpolated phase-shift at a given
value of $\omega$.  
The parameters in the right panel were determined by fitting the three 
lowest states of the spectrum to the perturbative expressions of 
sect.~\ref{sect:scatteringstates}. 
  }
\label{fig:extraction}
\end{figure}

By looking at different energy eigenvalues, the 
effective range parameters can be extracted
through the relation 
\begin{equation}\label{eqn:ERE}
p^{2l+1}\text{cot}\delta_l(p)=-1/a_l+1/2\ r_l p^2\ +\ .\ .\ .
\ \ =\ \   -1/a_l+1/2\ r_l m E\ +\ .\ .\ .\ ,
\end{equation} 
where $E$ is any relative energy eigenvalue which is low enough to ensure convergence 
of the ERE.
For example, the low-energy spectrum of the confined $^1S_0$ system 
can be used 
to extract the scattering length and effective range using eq.~(\ref{eq:NNHO}).
We show these extracted parameters at the 1-$\sigma$ level in
Fig.~(\ref{fig:extraction}).  
These  extracted parameters vary with $\omega^2$ in a way that is consistent
with expectations and converge to the exact result.
For a system with ERE parameters that 
are of natural size, the perturbative expressions from 
sect.~\ref{sect:scatteringstates} can also be used 
to extract these parameters.  
In Fig.~\ref{fig:extraction} the extracted `scattering volume', $a_1$, 
and `effective momentum', $r_1$, 
in the $^3P_0$ channel, determined through the 
perturbative expressions,
are shown.
The behavior as $\omega^2\rightarrow 0$ is consistent with the exact result.
%

\section{Conclusion}
\noindent
The NN phase-shift below the inelastic threshold can be determined
from the eigenvalue spectrum  of two interacting nucleons confined to move in a
harmonic oscillator potential.
The conventional discussions of scattering from a potential that falls faster
than $1/r$, and the connection between the scattering amplitude and the
location of poles in the complex energy-plane 
corresponding to bound-states
is complicated by the fact that
the harmonic potential is confining
and asymptotic scattering states cannot be defined 
for any non-zero value of the harmonic
oscillator frequency, $\omega$.    
As a result, the zero-range relation between the scattering phase-shift and the
energy-eigenvalues is modified by the non-zero value of the harmonic potential
within the range of the nuclear interaction, giving rise to finite-range corrections.  These corrections are not present when dealing with a pionless EFT description of two nucleons confined within a harmonic trap (when the cutoff is taken to infinity).   However, any pionful theory of the nuclear interaction that describes nuclear processes above the t-channel cut will have to address these finite-range issues.

We have studied these aspects numerically for two nucleons confined by a harmonic potential.
The nuclear interaction was modeled by the JISP16 potential, but our results are general and can be applied to other phenomenological or chiral effective field theory interactions. We have explored uncoupled channels and found that for small values of
$\omega$, the low-energy phase shift can be extracted from the
energy-eigenvalues through an extrapolation to $\omega=0$.  
At the level of precision to which  we have performed the calculations, the
energy-eigenvalues combined with the zero-range relation supplemented by an
extrapolation to  $\omega=0$ allow for the determination of the low-energy
NN elastic scattering phase shifts.  Further, such calculations
enable a precise determination of the deuteron binding energy.

Since the methods we present here are clearly non-perturbative and include all antisymmetrization effects,
an interesting application of eq.~(\ref{eq:NNHO}) would be to the elastic
scattering of two nuclear systems, with 
one or both composed of more than one nucleon, below inelastic and re-arrangement thresholds~\cite{Stetcu:2009ic}.  The processes we have in mind are 
$nd$, $nt$ and $n\alpha$ scattering.
Calculations of three-, four-, and five-nucleon systems can be 
performed within harmonic  potentials with small $\omega$ (to access low-energy phase shifts and minimize finite-range effects), and an application of 
eq.~(\ref{eq:NNHO}), modified by the reduced mass and an appropriate 
subtraction for the center-of-mass energy, 
and extrapolation to  $\omega=0$ would give the scattering phase-shift at 
low energies.  
This method contrasts with those currently in use, such as Faddeev~\cite{Witala:2004pv}, Faddeev-Yakubovsky~\cite{Lazauskas:2004uq}, AGS~\cite{Deltuva:2009fp}, Hyperspherical Harmonics~\cite{Marcucci:2009zz,Viviani:2008td}, 
 NCSM/RGM~\cite{Quaglioni:2009mn}, GFMC~\cite{Nollett:2006su}, and J-matrix methods~\cite{Shirokov:2008jv}.
There remains technical challenges to obtaining sufficient convergence with increasing $N_{MS}$ and/or to extending the corrections for finite $\omega$ to higher order terms.  A possible strategy to alleviate such issues is through the use of HO-based EFT methods~\cite{Stetcu:2010xq,Haxton:2007hx}.  Work in this direction is under way.

\begin{acknowledgments}
\noindent
TL and MJS thank the co-organizers of the INT workshop ``Simulations and Symmetries: Cold Atoms, LQCD, and Few-Hadron Systems", H. Hammer and D. Phillips, for
providing a stimulating environment in which part of this work was
accomplished. We thank S. Beane, A. Nicholson, S. Quaglioni, and I. Stetcu for their critical reading of this manuscript.  We also thank A. I. Mazur and A. M. Shirokov for
providing valuable scattering phase shifts for the JISP16 interaction
from solving the Schr\"odinger equation. The work of
TL was performed under the auspices of the U.S.~Department of Energy
by Lawrence Livermore National Laboratory under Contract
DE-AC52-07NA27344 and the UNEDF SciDAC grant DE-FC02-07ER41457. The work of MJS was supported in part by the
U.S.~Dept.~of Energy under Grant No.~DE-FG03-97ER4014. The work of JPV
was supported in part by the U.S.~Dept.~of Energy under Grant
No.~DE-FG02-87ER40371.  The work of AS was supported in part by the Natural Sciences and Engineering Research Council of Canada (NSERC) and the Helmholtz Alliance Program of the Helmholtz Association, contract HA216/EMMI ``Extremes of Density and Temperature:  Cosmic Matter in the Laboratory".  TRIUMF receives funding via a contribution through the National Research Council Canada.
\end{acknowledgments}

%
%

\end{document}